\documentclass[aps,prd,onecolumn,showpacs,showkeys,amsmath,amssymb,superscriptaddress,floatfix,nofootinbib]{revtex4}
\usepackage{amsfonts}
\usepackage{amsmath}
\usepackage{graphicx}
\usepackage{subfig}
\usepackage{dcolumn}
\usepackage{natbib}
\usepackage{bm}
\usepackage{booktabs}
\usepackage{slashed}

\usepackage{ulem}
\usepackage{color}

\makeatletter
\newcommand{\figcaption}{\def\@captype{figure}\caption}
\newcommand{\tabcaption}{\def\@captype{table}\caption}

\newcommand{\Rmnum}[1]{\expandafter\@slowromancap\romannumeral #1@}
\def\hlinewd#1{%
  \noalign{\ifnum0=`}\fi\hrule \@height #1 \futurelet
   \reserved@a\@xhline}
\makeatother
\def\dab{\int^{\alpha_{max}}_{\alpha_{min}}d\alpha\int^{\beta_{max}}_{\beta_{min}}d\beta}
\def\qq{\langle\bar qq\rangle}
\def\GGa{\langle GG\rangle}
\def\GGb{\langle \alpha_sGG\rangle}
\def\GGGa{\langle GGG\rangle}
\def\GGGb{\langle g_s^3fGGG\rangle}

\def\JJb{\langle g_s^4jj\rangle}
\def\f(s){[(\alpha+\beta)m^2-\alpha\beta s]}
\def\non{\\ \nonumber}
\begin{document}

\title{Masses of the Bottom-Charm  Hybrid $\bar bGc$ States}

\author{Wei Chen}
\email{wec053@mail.usask.ca}
\affiliation{Department of Physics
and Engineering Physics, University of Saskatchewan, Saskatoon, SK, S7N 5E2, Canada}
\author{T. G. Steele}
\email{tom.steele@usask.ca}
\affiliation{Department of Physics
and Engineering Physics, University of Saskatchewan, Saskatoon, SK, S7N 5E2, Canada}
\author{Shi-Lin Zhu} 
\email{zhusl@pku.edu.cn}
\affiliation{Department of Physics and State Key Laboratory of Nuclear Physics and Technology,\\
Peking University, Beijing 100871, China }

\begin{abstract}
QCD sum rules are used to study the mass spectrum of bottom-charm  hybrid $\bar bGc$ systems. The correlation functions and the 
spectral densities are calculated up to dimension six condensates at leading order of $\alpha_s$ for several $J^P$ quantum numbers. 
After performing the QCD sum rule analysis, we predict the masses of the $J^P=0^{-}, 0^{+}, 1^{-}, 1^{+}, 2^{-}, 2^{+}$ bottom-charm 
hybrids. These mass predictions show a similar supermultiplet structure as the bottomonium and charmonium  hybrids. Using the QCD 
sum-rule mass predictions we analyze the possible hadronic decay patterns of the $\bar cGc$, $\bar bGc$ and $\bar bGb$ hybrids including 
the open-flavour and hidden-flavour mechanisms.
\end{abstract}



\maketitle

%================================================================================
%================================================================================
\section{Introduction}\label{sec:INTRODUCTION}
%================================================================================
%================================================================================
The existence of hybrid mesons was suggested by Jaffe and Johnson in 1976~\cite{1976-Jaffe-p201-201}. Composed of a quark-antiquark 
pair and an excited gluonic field, hybrid mesons can carry exotic quantum numbers $J^{PC}=0^{--}, 0^{+-}, 1^{-+}, 2^{+-}$ {\it etc}. These exotic quantum numbers are not accessible for a quark-antiquark state, although hybrids can also have non-exotic quantum numbers and could in principle mix with quark-antiquark states. The observation of hybrids is one of the most important topics in hadronic physics, as evidenced by many experimental facilities such as PEPII, KEKB, BESIII, PANDA and LHCb that will search for hybrid mesons.

The spectrum of the light hybrid mesons was studied in several different approaches, such as the Bag model~\cite{1983-Barnes-p241-241, 1983-Chanowitz-p211-211}, the flux tube model~\cite{1985-Isgur-p869-869, 1995-Close-p1706-1709, 1995-Barnes-p5242-5256, 1999-Page-p34016-34016, 1995-Close-p233-254}, lattice QCD~\cite{1999-McNeile-p264-266, 1999-Lacock-p261-263, 2003-Bernard-p74505-74505, 
2005-Hedditch-p114507-114507} and QCD sum rules~\cite{1983-Govaerts-p262-262, 1984-Govaerts-p1-1, 1984-Latorre-p169-169, 1987-Latorre-p347-347, 1986-Balitsky-p265-273, 2003-Jin-p14025-14025, 2000-Chetyrkin-p145-150, 1999-Zhu-p97502-97502}. These approaches 
make very different mass predictions for the hybrids. To date, there has been some evidence of the exotic light hybrid with $J^{PC}=1^{-+}$~\cite{1997-Thompson-p1630-1633,1998-Abele-p175-184,1999-Abele-p349-355, 2007-Adams-p27-31} (see Refs.~\cite{2007-Klempt-p1-202, 2012-Ketzer-p25-25} for recent reviews).  Some studies exist for heavy quarkonium hybrids in the constituent gluon model~\cite{1978-Horn-p898-898}, the flux tube model~\cite{1995-Barnes-p5242-5256}, QCD sum rules~\cite{1985-Govaerts-p215-215,1985-Govaerts-p575-575,
1987-Govaerts-p674-674,1999-Zhu-p31501-31501,2012-Qiao-p15005-15005, 2012-Harnett-p125003-125003, 2012-Berg-p34002-34002}, 
nonrelativistic QCD~\cite{1998-Chiladze-p34013-34013} and lattice QCD~\cite{1990-Perantonis-p854-868, 1999-Juge-p4400-4403, 2006-Liu-p54510-54510, 
2006-Luo-p34502-34502, 2011-Liu-p140-140, 2012-Liu-p126-126}. 
However, to our knowledge the bottom-charm hybrids have not been studied in any of these methods.

In a recent paper~\cite{2013-Chen-p19-19}, we have studied the charmonium and bottomonium hybrids $\bar cGc$ and $\bar bGb$ using the QCD sum rule method. 
We consider the following operators which couple to the hybrid states with definite $J^{P(C)}$ quantum numbers 
\begin{eqnarray}
\nonumber
J^{(1)}_{\mu}&=&g_s\bar Q_1\frac{\lambda^a}{2}\gamma^{\nu}G^a_{\mu\nu}Q_2,~~~~~~~J^{PC}=1^{-(+)}, 0^{+(+)},
\\
J^{(2)}_{\mu}&=&g_s\bar Q_1\frac{\lambda^a}{2}\gamma^{\nu}\gamma_5G^a_{\mu\nu}Q_2,~~~~J^{PC}=1^{+(-)}, 0^{-(-)}, \label{currents}
\non
J^{(3)}_{\mu\nu}&=&g_s\bar Q_1\frac{\lambda^a}{2}\sigma_{\mu}^{\alpha}\gamma_5 G^a_{\alpha\nu}Q_2,~~~~J^{PC}=2^{-(+)}, 1^{+(+)}, 1^{-(+)}, 0^{-(+)}\,.
\end{eqnarray}
in which $Q_1$ and $Q_2$ are the heavy quark fields with masses $m_1$ and $m_2$, $g_s$ is the strong coupling constant, $\lambda^a$ 
are the Gell-Mann SU(3) matrices and $G^a_{\mu\nu}$ is the gluon field strength. It should be pointed out, however, the operators in Eq.~\eqref{currents} with $Q_1\neq Q_2$ 
carry no definite C-parities. The signs in the parentheses are the corresponding C-parities for the hidden flavour (equal mass) 
%\tcomm{would it be better to say flavour instead of mass?}\crep{The name ``equal mass hybrids'' was used in Govaerts' paper. I agree with you ``flavour'' sounds better. How about ``hidden-flavour''?} 
systems with $Q_1=Q_2$. By replacing $G^a_{\mu\nu}$ with $\tilde G^a_{\mu\nu}=\frac{1}{2}\epsilon_{\mu\nu\alpha\beta}G^{\alpha\beta,a}$, we can also obtain the corresponding operators $\tilde J^{(1)}_{\mu}, \tilde J^{(2)}_{\mu}, \tilde J^{(3)}_{\mu\nu}$ with opposite parities. These hybrid interpolating currents were originally studied to calculate the masses of the hidden flavour $\bar cGc$ and $\bar bGb$ hybrids for $Q_1=Q_2$ in Refs.~\cite{1985-Govaerts-p215-215,1987-Govaerts-p674-674} and the open flavour heavy-light $\bar QGq$ hybrids 
for $Q_1\neq Q_2$ in Ref.~\cite{1985-Govaerts-p575-575}. However, only the perturbative and dimension four gluon condensate contributions were calculated for the correlation functions in these papers, which resulted in unstable hybrid sum rules and hence unreliable mass predictions for some channels. Recently, dimension-six condensates have been shown to stabilize the sum-rule mass predictions of the $J^{PC}=1^{--}$~\cite{2012-Qiao-p15005-15005}, $1^{++}$~\cite{2012-Harnett-p125003-125003} and $0^{-+}$~\cite{2012-Berg-p34002-34002} channels. The dimension six contributions are thus very important because they stabilize the hybrid sum rules.

In Ref.~\cite{2013-Chen-p19-19}, we re-analyzed all the channels with $J^{PC}=0^{-+}, 0^{--}, 0^{++}, 0^{+-}, 1^{-+}, 1^{--}, 1^{+-}, 
1^{++},  2^{-+},  2^{++}$ by including the tri-gluon condensate contributions and updated the mass spectrum of $\bar cGc$ and 
$\bar bGb$ hybrids, confirming the supermultiplet structures of the heavy quarkonium hybrid spectrum found in lattice 
QCD~\cite{2012-Liu-p126-126} and the P-wave quasigluon approach~\cite{2008-Guo-p56003-56003}.

In this paper, we extend our investigation to $\bar bGc$ ($\bar cGb$) systems using the interpolating currents in Eq.~\eqref{currents} with $Q_1\neq Q_2$. 
%\trem{These systems are more complex because they contain two different flavours of heavy quarks and hence there are multiple energy scales involved in the loop calculations.} 
%\trem{To our knowledge, these $\bar bGc$ hybrids have not been studied before.}\crep{I remove this sentence because it is almost the same with the last sentence of the sencond paragraph.} 
In this situation, the $\bar bGc$ type of currents in Eq.~\eqref{currents} couple to the charged hybrid states with no definite C-parities. This is very similar to the heavy-light hybrids studied in QCD sum rules~\cite{1985-Govaerts-p575-575} and lattice QCD~\cite{2013-Moir-p21-21}. With their 
special hadronic configuration, the mass prediction of the bottom-charm $\bar bGc$ hybrids can provide important information for future experimental searches.

For the $\bar bGc$ currents in Eq.~\eqref{currents}, $J^{(1)}_{\mu}$ and $\tilde J^{(2)}_{\mu}$ both couple to the $J^P=1^{-}, 0^{+}$ states while $J^{(2)}_{\mu}$ and $\tilde J^{(1)}_{\mu}$ both couple to the $J^P=1^{+}, 0^{-}$ states. Although multiple operators exist for a given quantum number, they may have different couplings to the ground and excited states, so 
one should not necessarily expect the same mass predictions.  Generally, the operator leading to the smallest mass prediction would provide the best determination of the ground state.
For the spin-2 states, the tensor currents $J^{(3)}_{\mu\nu}$ and $\tilde J^{(3)}_{\mu\nu}$ couple to $J^P=2^-$ and $J^P=2^+$ channels respectively. We will calculate the correlation functions up to dimension six tri-gluon condensate contributions to perform the QCD sum rule analysis. 

The paper is organized as follows. In Sec.~\Rmnum{2}, we calculate the $\bar bGc$ correlation functions and spectral densities using the 
hybrid interpolating currents with various quantum numbers in Eq.~\eqref{currents}. In Sec.~\Rmnum{3}, we perform the numerical 
analysis and extract masses of the $\bar bGc$ hybrid states. We study the possible decay patterns of the $\bar cGc$, $\bar bGc$ 
and $\bar bGb$ hybrids in Sec.~\Rmnum{4}. The last section is a brief summary.

%================================================================================
%================================================================================
\section{QCD Sum Rule and Spectral Densities}\label{sec:QSR}
%================================================================================
%================================================================================
In the past few decades, QCD sum rules have been widely used to study hadronic structures~\cite{1979-Shifman-p385-447,
1985-Reinders-p1-1,2000-Colangelo-p1495-1576}. We consider the two-point correlation function:
\begin{eqnarray}
\Pi_{\mu\nu}(q)= i\int
d^4xe^{iq\cdot x}\langle0|T[J_{\mu}(x)J_{\nu}^{\dag}(0)]|0\rangle, \label{equ:Pi}
\end{eqnarray}
where $J_{\mu}$ is the hybrid interpolating current in Eq.~\eqref{currents}. Since these currents are not conserved, 
the two-point correlation functions have the following structures:
\begin{eqnarray}
i\int d^4x \,e^{iq\cdot x}\,\langle 0|T [J_\mu(x)J_\nu^\dagger(0)]|0\rangle
&=&\left[\frac{q_\mu q_\nu}{q^2}-g_{\mu\nu}\right]\Pi_V(q^2)+\frac{q_\mu q_\nu}{q^2}\Pi_S(q^2), \label{pivector}\\
i\int d^4x\,e^{iq\cdot x}\,\langle0|T[J_{\mu\nu}(x)J_{\rho\sigma}^{\dag}(0)]|0\rangle
&=&\left[\eta_{\mu\rho}\eta_{\nu\sigma}+\eta_{\mu\sigma}\eta_{\nu\rho}
-\frac{2}{3}\eta_{\mu\nu}\eta_{\rho\sigma}\right]\Pi_T(q^2)+\ldots \,, \label{pitensor}
\end{eqnarray}
where $\eta_{\mu\nu}=q_{\mu}q_{\nu}/q^2-g_{\mu\nu}$. The imaginary parts of the invariant functions $\Pi_V(q^2)$, 
$\Pi_S(q^2)$ and $\Pi_T(q^2)$ refer to pure spin-1, spin-0 and spin-2 intermediate states, respectively. 
In Eq.~(\ref{pitensor}), the invariant structures for spin-0 and spin-1 are not written out explicitly because we will not 
consider contributions arising from these terms in this paper.

The correlation function can be described at both the hadron level and the quark-gluon level.
To determine the correlation function at the hadron level, we use the dispersion relation
\begin{eqnarray}
\Pi(q^2)=\frac{(q^2)^N}{\pi}\int_{(m_1+m_2)^2}^{\infty}\frac{\mbox{Im}\Pi(s)}{s^N(s-q^2-i\epsilon)}ds+\sum_{n=0}^{N-1}b_n(q^2)^n, 
\label{dispersion relation}
\end{eqnarray}
where 
\begin{eqnarray}
\rho(s)\equiv\frac{1}{\pi}\mbox{Im}\Pi(s)=\sum_n\delta(s-m_n^2)\langle0|J_{\mu}|n\rangle\langle n|J_{\mu}^{\dagger}|0\rangle
=f_X^2m_X^8\delta(s-m_X^2)+ \mbox{continuum},  \label{Phenrho}
\end{eqnarray}
in which we use a narrow resonance approximation and write the spectral function $\rho(s)$ (the imaginary part of $\Pi(s)$) 
as a sum over a series of zero-width $\delta$ functions. Finally, the pole plus continuum approximation is adopted to pick out 
the lowest lying resonance. 
The unknown subtraction constants 
$b_n$ in the right hand side of Eq.~\eqref{dispersion relation} can be removed by taking the Borel transform of $\Pi(q^2)$. 
The intermediate states $\vert n\rangle$ must have the same quantum numbers as the interpolating currents 
$J_{\mu}$. In Eq.~\eqref{Phenrho}, $m_X$ denotes the mass of the lowest lying resonance, and the dimensionless quantity $f_X$ 
is the coupling of the resonance to the current
\begin{eqnarray}
\langle0|J_{\mu}|X\rangle&=&f_Xm_X^4\epsilon_{\mu}\,, \\
\langle0|J_{\mu\nu}|X\rangle&=&f_Xm_X^4\epsilon_{\mu\nu}\,,
\end{eqnarray}
in which $\epsilon_{\mu}$ and $\epsilon_{\mu\nu}$ are the (spin-1) polarization vector and (spin-2) polarization tensor.

To evaluate the correlation function $\Pi(q^2)$ at the quark-gluon level, we first need to determine the full quark propagator. 
For $\bar bGc$ hybrid systems, the quark condensates and quark-gluon mixed condensates are expressed in terms of the gluon condensate 
and tri-gluon condensate via the heavy quark mass expansion and hence give no contributions to the correlation function.
Taking into account only the gluon condensate and tri-gluon condensate contributions, we use the full quark propagator in 
momentum space~\cite{1985-Reinders-p1-1}
\begin{eqnarray}
iS_{ab}(p) &=&
\frac{i\delta_{ab}}{\slashed{p}-m}+\frac{i}{4}g_s\frac{\lambda^n_{ab}}{2}G_{\mu\nu}^n
\frac{\sigma^{\mu\nu}(\slashed{p}+m)+(\slashed{p}+m)\sigma^{\mu\nu}}{(p^2-m^2)^2},
\end{eqnarray}
where $\sigma^{\mu\nu}=\frac{i}{2}\left[\gamma^\mu\,,\gamma^\nu\right]$, and $a\,,b$ are color indices. To calculate the Wilson coefficients 
at leading order in $\alpha_s$, the perturbative, gluon condensate $\GGb$ and tri-gluon condensate $\GGGb$ contributions are represented 
in Fig.~\ref{fig1}, Fig.~\ref{fig2} and Fig.~\ref{fig3}, respectively. In Fig.~\ref{fig3}a, the condensate $\langle DDG\rangle$ can be expressed 
in terms of $\GGGb$ and $\JJb$ by using the equation of motion, Bianchi identities and commutation relations. Within the vacuum factorization 
assumption, the contribution of $\JJb$ is proportional to the square of $g_s^2\qq$ and thus will be neglected in this work since it is a small numerical effect compared to the gluonic condensates~\cite{2013-Chen-p19-19}. 
%%%%%%%%%%%%%%%%%%%%%%%%%%%%%%%%%%%%%%%%%%%%%%%%%%%%%%%%%%%%%%%%%%%%%%
%\begin{figure}
\begin{center}
\includegraphics[scale=0.6]{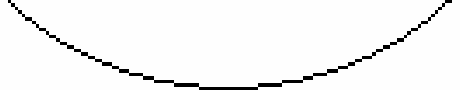}
\figcaption{Feynman diagram representing the perturbative contribution to the correlation functions. Solid and curly lines represent quark 
and gluon propagators respectively, while the dashed line represents the interpolating current.} \label{fig1}
\end{center}
%\end{figure}
%%%%%%%%%%%%%%%%%%%%%%%%%%%%%%%%%%%%%%%%%%%%%%%%%%%%%%%%%%%%%%%%%%%%%
%%%%%%%%%%%%%%%%%%%%%%%%%%%%%%%%%%%%%%%%%%%%%%%%%%%%%%%%%%%%%%%%%%%%%%%
%\begin{figure}
\begin{center}
\includegraphics[scale=0.6]{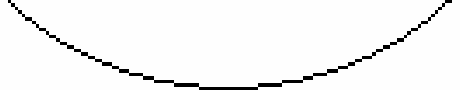}
\figcaption{Feynman diagram representing the $\langle\alpha_sGG\rangle$
contribution to the correlation functions.} \label{fig2}
\end{center}
%\end{figure}
%%%%%%%%%%%%%%%%%%%%%%%%%%%%%%%%%%%%%%%%%%%%%%%%%%%%%%%%%%%%%%%%%%%%%%%%%%%%%%
%%%%%%%%%%%%%%%%%%%%%%%%%%%%%%%%%%%%%%%%%%%%%%%%%%%%%%%%%%%%%%%%%%%%%%
%\begin{figure}
\begin{center}
\includegraphics[scale=0.6]{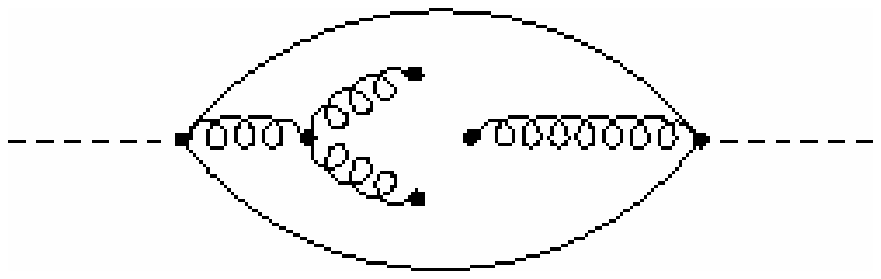}
 \includegraphics[scale=0.6]{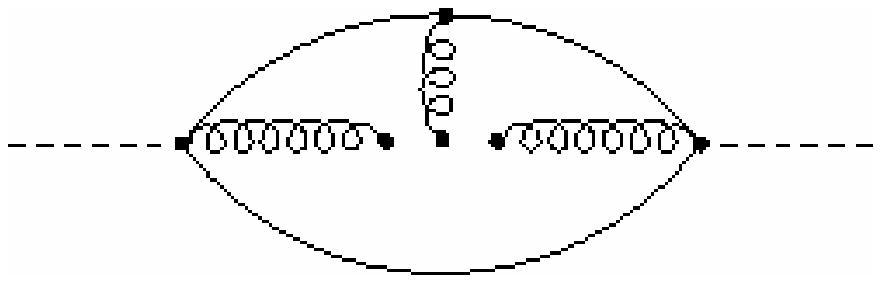}
 \centerline{\hspace{-0.04in} {(a)} \hspace{1.90in}{ (b)}}
 \figcaption{Feynman diagrams representing the $\langle DDG\rangle$ (a) and $\GGGb$ (b)
contributions to the correlation functions.
} \label{fig3}
\end{center}
%\end{figure}
%%%%%%%%%%%%%%%%%%%%%%%%%%%%%%%%%%%%%%%%%%%%%%%%%%%%%%%%%%%%%%%%%%%%

Appendix~\ref{sec:rhos} presents the results for the correlation functions and the spectral densities up to dimension six condensate contributions. From these expressions, we find that the perturbative contributions are invariant and the gluon condensate and tri-gluon condensate contributions change sign under 
the replacement $G^a_{\mu\nu}\to\tilde G^a_{\mu\nu}$ of the interpolating currents. For the same spin channels (spin-0 or spin-1), the specral 
densities from $J^{(1)}_{\mu}$ are identical to those from $J^{(2)}_{\mu}$ except for the additional minus signs for the odd power of the 
$m_1m_2$ proportional terms. For these formulae, it is straightforward to check that in the equal mass limit one recovers known results in 
Ref.~\cite{2013-Chen-p19-19}. Note the presence of $\delta(s-\tilde m^2)$ and its derivatives in the tri-gluon condensate contributions~\eqref{tri-gluon 
condensate contribution} to compensate for the singular behavior of the spectral densities at threshold $s=(m_1+m_2)^2$. 

We can establish a sum rule for the hybrid mass by comparing the correlation function calculated at the quark-gluon level with those from 
the dispersion relation at the hadron level. The Borel transform is applied at both the quark-gluon and hadron levels to pick out the lowest 
lying resonance, eliminate the unknown subtraction constants in Eq.~\eqref{dispersion relation}, and enhance the operator-product expansion (OPE) convergence. 
Using the spectral function in Eq.~(\ref{Phenrho}), we arrive at
\begin{eqnarray}
f_X^2 m_X^{8+2k}e^{-m_X^2/M_B^2}=\int_{(m_1+m_2)^2}^{s_0}ds\,s^k\,\rho(s)\,e^{-s/M_B^2}=\mathcal{L}_{k}\left(s_0, M_B^2\right)\,,
\label{sumrule}
\end{eqnarray}
where $s_0$ is the continuum threshold parameter and $M_B$ is the Borel mass. Then we extract the hybrid mass via
\begin{eqnarray}
m_X^2=\frac{\mathcal{L}_{1}\left(s_0\,, M_B^2\right)}{\mathcal{L}_{0}\left(s_0\,, M_B^2\right)}\,, \label{mass}
\end{eqnarray}
and the corresponding coupling constant via
\begin{eqnarray}
f_X^2=\frac{\mathcal{L}_{0}\left(s_0, M_B^2\right)}{m_X^8}e^{m_X^2/M_B^2}, \label{coupling}
\end{eqnarray}
in which $m_X$ denotes the $\bar bGc$ hybrid mass as defined in Eq.~\eqref{mass}.

%================================================================================
%================================================================================
\section{Numerical Analysis}\label{sec:NA}
%================================================================================
%================================================================================
We perform the QCD sum rule numerical analysis using the following values of the heavy quark masses
and the condensates~\cite{2009-Chetyrkin-p74010-74010, 2012-Narison-p259-263, 2010-Narison-p559-559,
2007-Kuhn-p192-215}:
\begin{eqnarray}
\nonumber &&m_c(\mu=m_c)=\overline m_c=(1.28\pm 0.02)~\mbox{GeV},
\non &&m_b(\mu=m_b)=\overline m_b=(4.17\pm 0.02)~\mbox{GeV},
\non &&\GGb=(7.5\pm2.0)\times 10^{-2}~\mbox{GeV}^4,
\\&& \GGGb=-(8.2\pm1.0)~\mbox{GeV}^2\GGb,
\label{qcd_parameters}
\end{eqnarray}
in which the charm and bottom quark masses are the running masses in the $\overline{\rm MS}$ scheme. 
The definition of the coupling constant $g_s$ has a minus sign difference in this work.
Furthermore, we take into account the scale dependence of these $\overline{\rm MS}$ masses in the leading order:
\begin{eqnarray}
m_c(\mu)=\overline m_c\bigg(\frac{\alpha_s(\mu)}{\alpha_s(\overline m_c)}\bigg)^{12/25},
\\m_b(\mu)=\overline m_b\bigg(\frac{\alpha_s(\mu)}{\alpha_s(\overline m_b)}\bigg)^{12/23}\, ,
\end{eqnarray}
where
\begin{eqnarray}
\alpha_s(\mu)&=&\frac{\alpha_s(M_{\tau})}{1+\frac{25\alpha_s(M_{\tau})}{12\pi}\log(\frac{\mu^2}{M_{\tau}^2})}, 
\quad \alpha_s(M_{\tau})=0.33, \label{alpha_cc}
\end{eqnarray}
is determined by evolution from the $\tau$ mass using Particle Data Group values~\cite{2012-Beringer-p10001-10001}. 
In the bottom-charm hybrid systems, there is a typical scale $\mu=\frac{\overline m_c+\overline m_b}{2}=2.73$ GeV which will be 
adopted in our sum rule analysis. 

There are two very important parameters in the sum rules Eq.~\eqref{sumrule}: the continuum threshold parameter $s_0$ and Borel 
mass $M_B$. To establish a reliable sum rule extracting the hybrid mass from Eq.~\eqref{mass}, one should obtain suitable working 
regions of these two parameters. In our analysis, we choose the value of $s_0$ around which the variation of the hybrid mass $m_X$ 
with $M_B^2$ is minimum. The Borel window is determined by the convergence of the OPE series and the pole contribution. The lower 
bound on $M_B^2$ is determined by imposing that the gluon condensate contribution is less than one fourth of the perturbative 
contribution while the tri-gluon condensate contribution is less than one fourth of the gluon condensate contribution. Requiring 
the pole contribution to be larger than $50\%$, we obtain the upper bound on $M_B^2$. 
%%%%%%%%%%%%%%%%%%%%%%%%%%%%%%%%%%%%%%%%%%%%%%%%%%%%%%%%%%%%%%%%%%%%%%
%\begin{figure}
\begin{center}
\begin{tabular}{c}
\scalebox{0.8}{\includegraphics{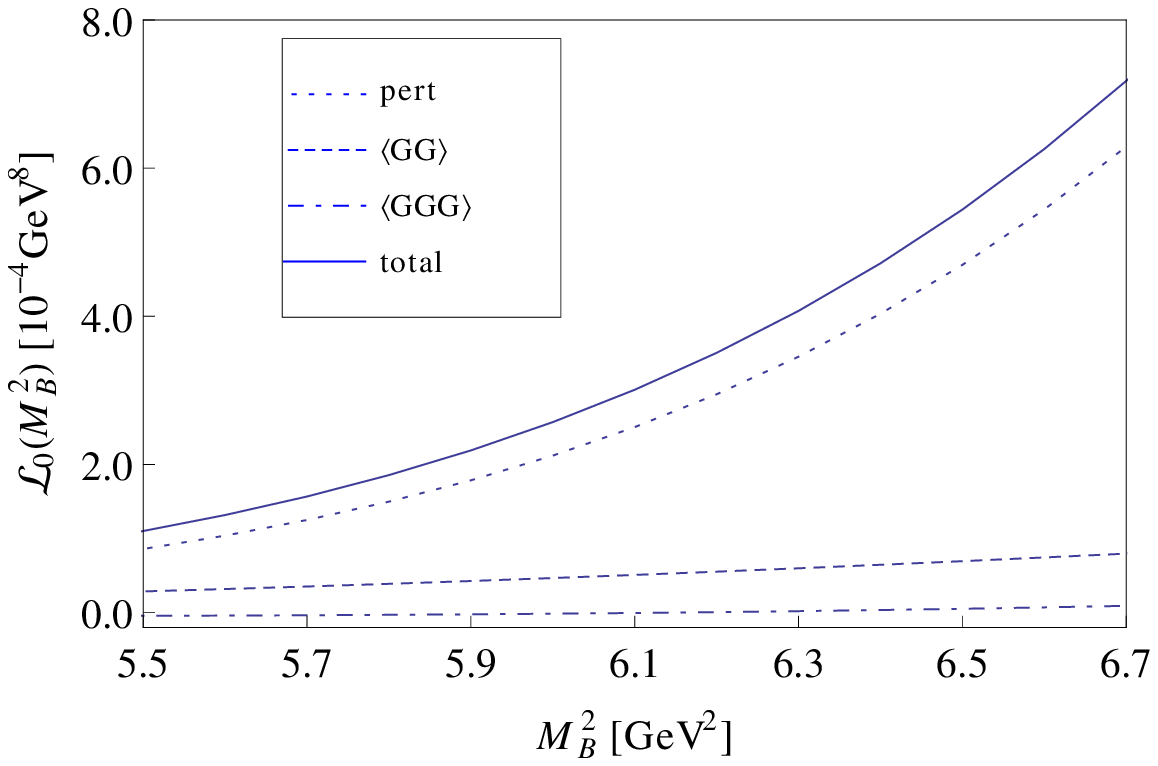}}
\end{tabular}
\figcaption{The contributions of each term in OPE series, including the perturbative term, gluon condensate $\GGb$ and tri-gluon condensate $\GGGb$ 
for the $\bar bGc$ hybrid with $J^P=1^{-}$ by using $\tilde J^{(2)}_{\mu}$.} \label{figOPEbc1-}
\end{center}
%\end{figure}
%%%%%%%%%%%%%%%%%%%%%%%%%%%%%%%%%%%%%%%%%%%%%%%%%%%%%%%%%%%%%%%%%%%%%%

The definition of the pole contribution (PC) is 
\begin{eqnarray}
\mbox{PC}(s_0, M_B^2)=\frac{\mathcal{L}_{0}\left(s_0\,, M_B^2\right)}{\mathcal{L}_{0}\left(\infty\,, M_B^2\right)}. \label{PC}
\end{eqnarray}

We first study the $J^P=1^{-}$ $\bar bGc$ hybrid by using the interpolating current $\tilde J^{(2)}_{\mu}$. 
The dominant non-perturbative 
contribution is the gluon condensate. After studying the OPE series, we show the OPE convergence for this channel in 
Fig.~\ref{figOPEbc1-}, from which we determine the lower bound on the Borel mass as $M^2_{\mbox{min}}=5.9$ GeV$^2$.
In Fig.~\ref{fig1-1}, we show the variations of the hybrid mass $m_X$ with the threshold parameter $s_0$ 
and the Borel mass $M_B^2$ for this channel. From the left plot, we choose the threshold parameter $s_0=52$ GeV$^2$ around which the 
variation of $m_X$ with $M_B^2$ is very weak, as shown in the right plot of Fig.~\ref{fig1-1}. Considering the constraint of the pole 
contribution, we obtain the upper bound of the Borel mass $M^2_{\mbox{max}}=6.5$ GeV$^2$. 
In this Borel window, we extract the mass for the $J^P=1^{-}$ $\bar bGc$ hybrid
\begin{eqnarray}
m_X=6.83\pm0.08\pm0.01\pm0.07~\text{GeV},
\end{eqnarray}
in which the errors come respectively from the continuum threshold $s_0$, the heavy quark masses $m_c, m_b$ and the gluon condensates $\GGb, \GGGb$. 
The error from the Borel mass $M_B$ is negligible since the mass sum rules is very stable in the Borel window (See Fig.~\ref{fig1-1} and Fig.~\ref{fig1-2}). 

We can also explore the $1^{-}$ $\bar bGc$ hybrid using the interpolating current $J^{(1)}_{\mu}$ in Eq.~\eqref{currents}. 
To obtain a significant Borel window for this channel, we relax the constraint by requiring the pole contribution be larger than $20\%$. This requirement of 
the pole contribution also occurs in two $0^-$ channels, $2^-$ and $2^+$ channels, $0^+$ channel for the current $\tilde J^{(2)}_{\mu}$.
Performing the same analysis as done above, we obtain the Borel window $9.00\,{\rm GeV^2} \leq M_B^2\leq 10.7$ GeV$^2$ 
while $s_0=62$ GeV$^2$ for $J^{(1)}_{\mu}$ with $J^P=1^-$. In contrast, we extract the hybrid mass $m_X=6.95\pm0.13\pm0.01\pm0.08$ GeV, which is $0.12$ GeV lower than 
the $1^{-}$ hybrid mass extracted from $\tilde J^{(2)}_{\mu}$. The Borel curves are shown in Fig.~\ref{fig1-2} and the numerical results in Table~\ref{table1}. 

%%%%%%%%%%%%%%%%%%%%%%%%%%%%%%%%%%%%%%%%%%%%%%%%%%%%%%%%%%%%%%%%%%%%%%
%\begin{figure}
\begin{center}
\begin{tabular}{lr}
\scalebox{0.65}{\includegraphics{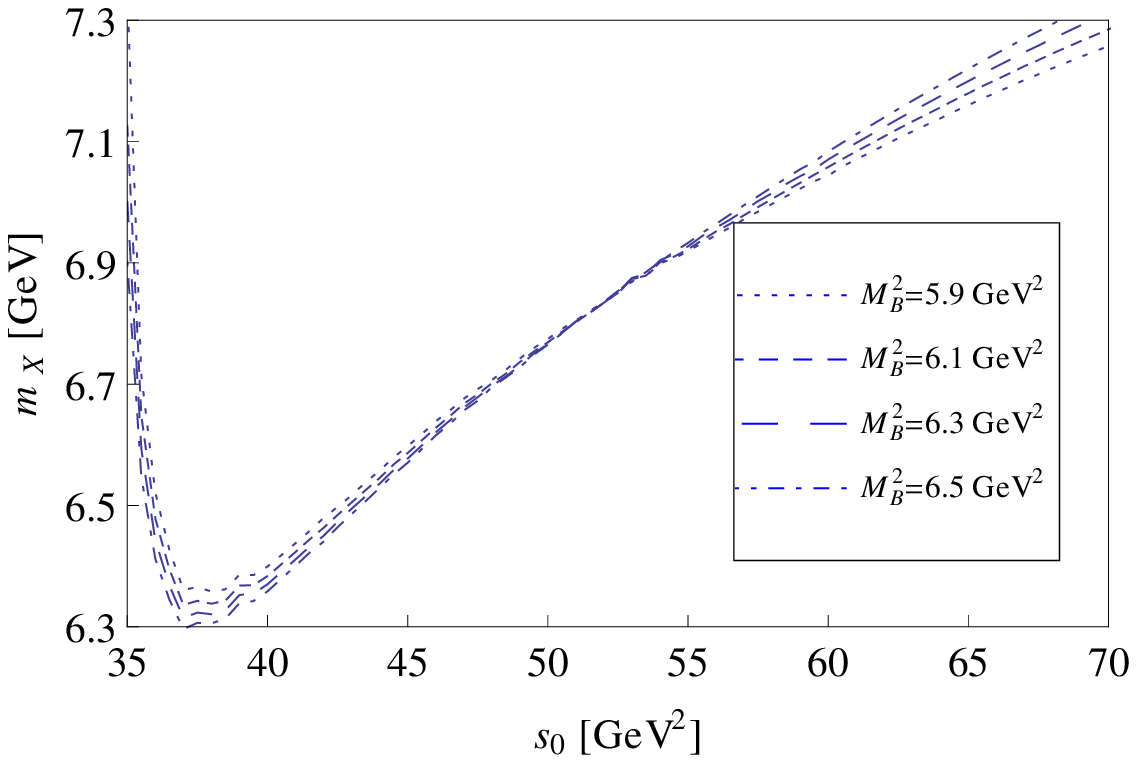}}&
\scalebox{0.65}{\includegraphics{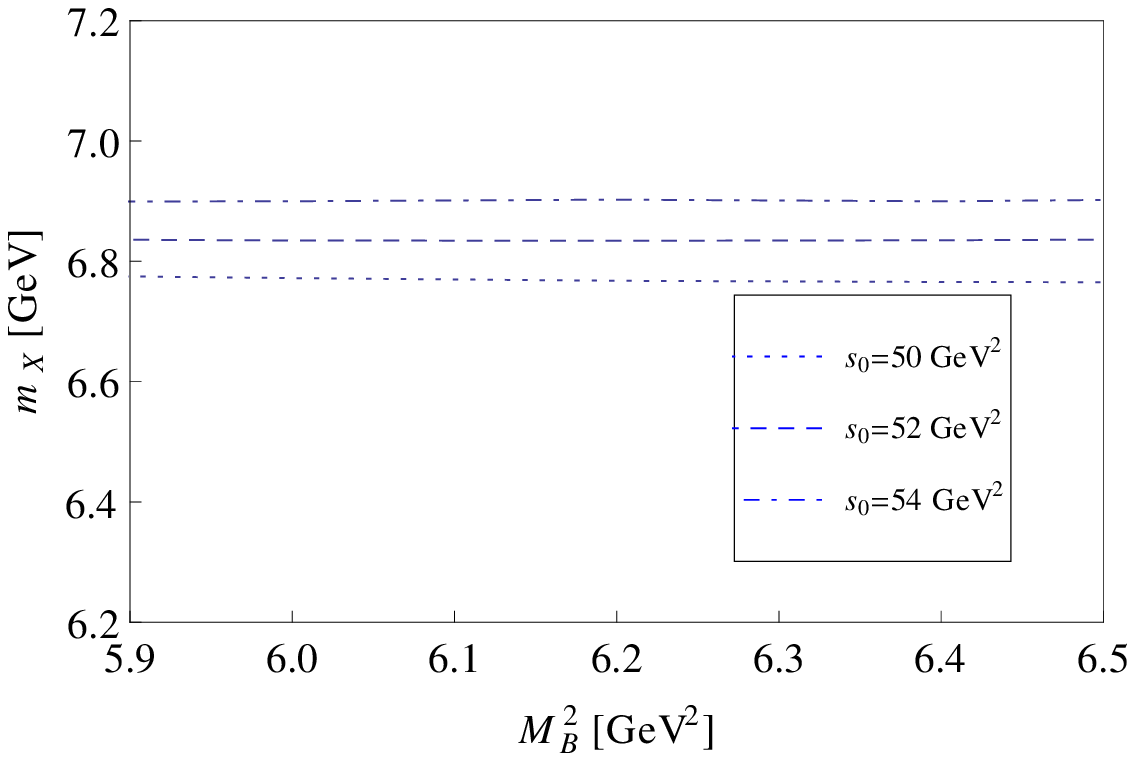}}
\end{tabular}
\figcaption{The variations of the hybrid mass $m_X$ with $s_0$ and $M_B^2$ for the $J^P=1^-$ channel by using $\tilde J^{(2)}_{\mu}$.} \label{fig1-1}
\end{center}
%\end{figure}
%%%%%%%%%%%%%%%%%%%%%%%%%%%%%%%%%%%%%%%%%%%%%%%%%%%%%%%%%%%%%%%%%%%%%%
%\begin{figure}
\begin{center}
\begin{tabular}{lr}
\scalebox{0.65}{\includegraphics{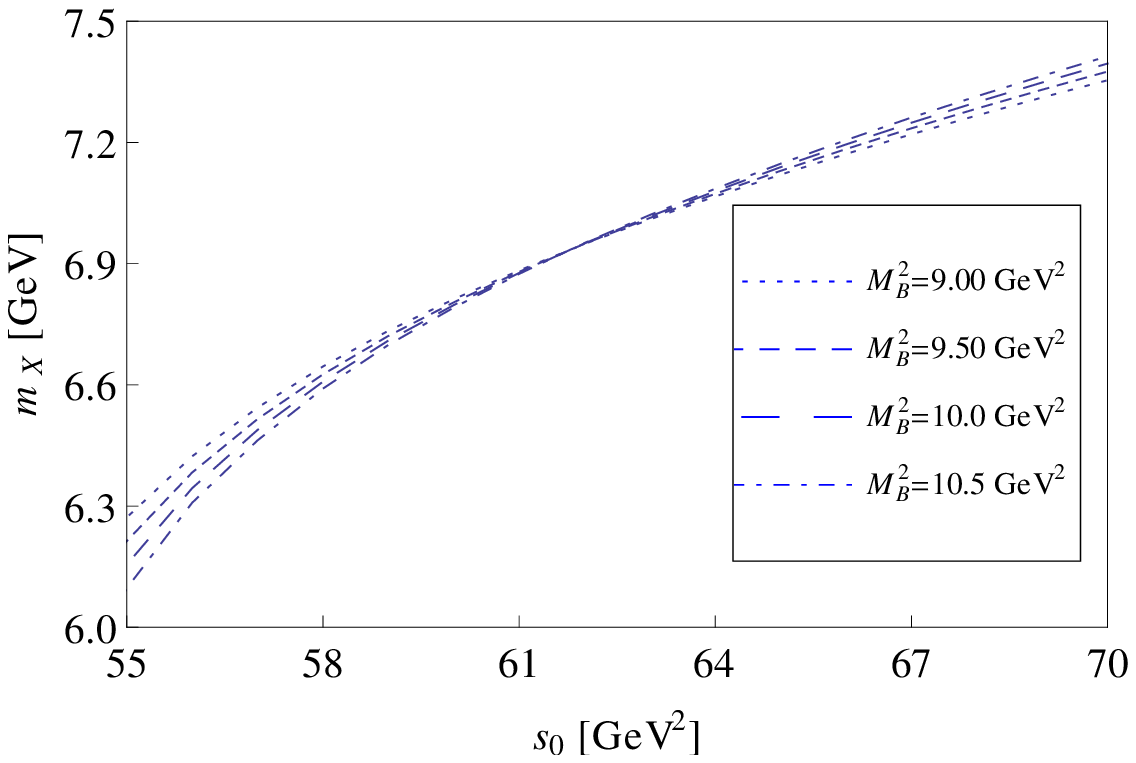}}&
\scalebox{0.65}{\includegraphics{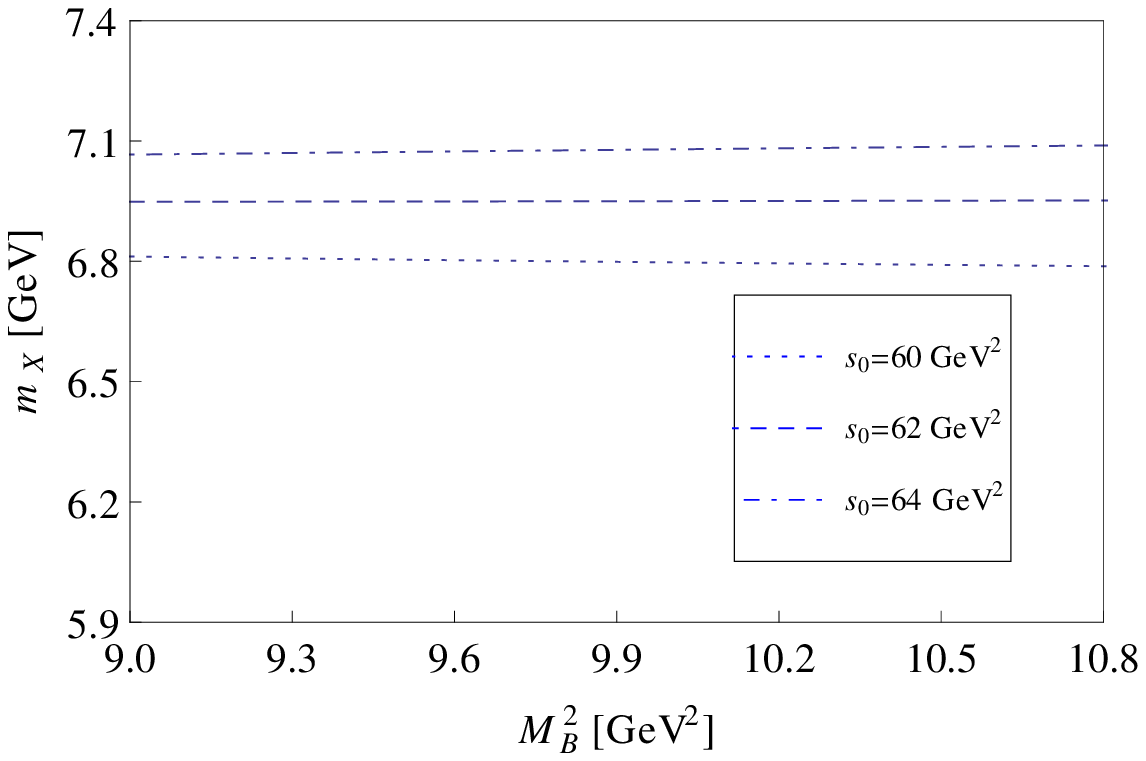}}
\end{tabular}
\figcaption{The variations of the hybrid mass $m_X$ with $s_0$ and $M_B^2$ for the $J^P=1^-$ channel by using $J^{(1)}_{\mu}$.} \label{fig1-2}
\end{center}
%\end{figure}

Using the spectral densities listed in Appendix~\ref{sec:rhos}, we perform the sum rule analyses for the other $\bar bGc$ hybrids with 
$J^P=0^-, 0^+, 1^+, 2^-, 2^+$. All these channels are stable enough to choose the suitable Borel windows among which one can establish 
reliable sum rules to extract the hybrid masses. We collect the numerical results including the hybrid masses, the threshold values, 
the Borel windows and the pole contributions in Table~\ref{table1}. In  cases where there exist multiple currents for the same quantum 
numbers, differences in the mass predictions imply that the currents differ in their relative couplings to the ground and excited state. 
The lowest mass prediction should be interpreted as the best determination of the ground state. As mentioned above,  
the errors of mass predictions come from the uncertainties in $s_0$, $m_c$ and $m_b$, $\GGb$ and $\GGGb$ respectively. 
One notes that both the errors from $s_0$ and the condensates 
$\GGb, \GGGb$ are important for mass predictions of two $1^-$ channels, lower $0^-$ and $0^+$ channels and $2^-$ channel. For the other 
channels, however, the main errors are from the uncertanties of QCD condensates. The errors from the heavy quark masses are very small for all 
channels.

%%%%%%%%%%%%%%%%%%%%%%%%%%%%%%%%%%%%%%%%%%%%%%%%%%%%%%%%%%%%%%%%%%%%
\begin{center}
\renewcommand{\arraystretch}{1.3}
\begin{tabular*}{16.5cm}{cccccccc}
%\begin{tabular}{ccccccc}
\hlinewd{.8pt}
& Operator & $J^{P}$     & $s_0(\mbox{GeV}^2)$&$[M^2_{\mbox{min}}$,$M^2_{\mbox{max}}](\mbox{GeV}^2)$&$m_X$\mbox{(GeV)}&PC(\%)& $f_X^2 (10^{-7})$\\
\hline
&$\tilde J^{(2)}_{\mu}$   & $1^{-}$   & 52  & $5.90\sim6.50 $ & $6.83\pm0.08\pm0.01\pm0.07$& 55.9 & $0.77\pm0.08\pm0.05\pm0.17$\\
&$\tilde J^{(1)}_{\mu}$   & $0^{-}$   & 61  & $10.4\sim11.5 $ & $6.90\pm0.12\pm0.01\pm0.09$& 23.4 & $2.21\pm0.42\pm0.16\pm0.22$\\
&$J^{(1)}_{\mu}$          & $1^{-}$   & 62  & $9.00\sim10.7 $ & $6.95\pm0.13\pm0.01\pm0.08$& 26.1 & $1.74\pm0.37\pm0.14\pm0.49$\\
&$J^{(3)}_{\mu\nu}$       & $2^{-}$   & 59  & $8.00\sim10.5 $ & $7.15\pm0.08\pm0.05\pm0.09$& 29.4 & $0.43\pm0.07\pm0.01\pm0.10$
\vspace{5pt}\\
&$\tilde J^{(2)}_{\mu}$   & $0^{+}$   & 69  & $10.9\sim12.0 $ & $7.37\pm0.12\pm0.07\pm0.12$& 22.8 & $0.87\pm0.17\pm0.01\pm0.23$\\
&$\tilde J^{(3)}_{\mu\nu}$& $2^{+}$   & 66  & $8.00\sim11.2 $ & $7.67\pm0.07\pm0.02\pm0.09$& 39.8 & $0.51\pm0.10\pm0.04\pm0.04$\\
&$J^{(2)}_{\mu}$          & $1^{+}$   & 71  & $5.90\sim8.00 $ & $7.77\pm0.06\pm0.05\pm0.13$& 59.4 & $2.19\pm0.34\pm0.02\pm0.57$\\
&$\tilde J^{(1)}_{\mu}$   & $1^{+}$   & 77  & $9.00\sim10.0 $ & $8.28\pm0.05\pm0.02\pm0.31$& 54.3 & $4.99\pm0.52\pm0.03\pm0.88$\\
&$J^{(1)}_{\mu}$          & $0^{+}$   & 84  & $10.4\sim12.2 $ & $8.55\pm0.05\pm0.02\pm0.37$& 55.7 & $7.14\pm0.62\pm0.02\pm0.86$
\vspace{5pt}\\
&$J^{(2)}_{\mu}$          & $0^{-}$   & 76  & $10.9\sim14.2 $ & $8.48\pm0.04\pm0.04\pm0.59$& 24.4 & $1.29\pm0.16\pm0.02\pm0.34$\\
\hline
\hlinewd{.8pt}
%\end{tabular}
\end{tabular*}
\tabcaption{Masses of the $\bar bGc$ hybrid states and the corresponding $s_0$,
Borel windows, pole contributions and operator-state overlap.\label{table1}}
\end{center}
%%%%%%%%%%%%%%%%%%%%%%%%%%%%%%%%%%%%%%%%%%%%%%%%%%%%%%%%%%%%%%%

To study the sensitivity of the hybrid sum rules to the values of gluon and tri-gluon condensates, we reanalyze the $\tilde J^{(2)}_{\mu}$ channel with $J^P=1^-$ by 
adopting another set of condensate values from Ioffe's recent review~\cite{2006-Ioffe-p232-277}: $\GGb=(0.016\pm0.013)$ GeV$^4$, $\GGGb=-4.69\GGb$ GeV$^2$=$-(0.075\pm0.061)$ 
GeV$^6$. For the current $\tilde J^{(2)}_{\mu}$ with $J^P=1^-$, the working regions of the hybrid sum rule with the above condensate values are $s_0=51$ GeV$^2$ and $4.1$ 
GeV$^2$ $\leq M_B^2\leq 4.7$ GeV$^2$. This Borel window is quite different with that obtained in Table~\ref{table1} while the continuum threshold $s_0$ is very similar. 
We show the variations of the hybrid mass with $s_0$ and $M_B^2$ for $\tilde J^{(2)}_{\mu}$ channel with $J^P=1^-$ in Fig.~\ref{fig1-Ioffe}. The hybrid mass is extracted 
as $m_X=6.69$ GeV, which is slightly lower than that obtained in Table~\ref{table1} using Narison's condensate values  \cite{2012-Narison-p259-263,2010-Narison-p559-559}.
%%%%%
%%%%%%%%%%%%%%%%%%%%%%%%%%%%%%%%%%%%%%%%%%%%%%%%%%%%%%%%%%%%%%%%%%%%%%
%\begin{figure}
\begin{center}
\begin{tabular}{lr}
\scalebox{0.65}{\includegraphics{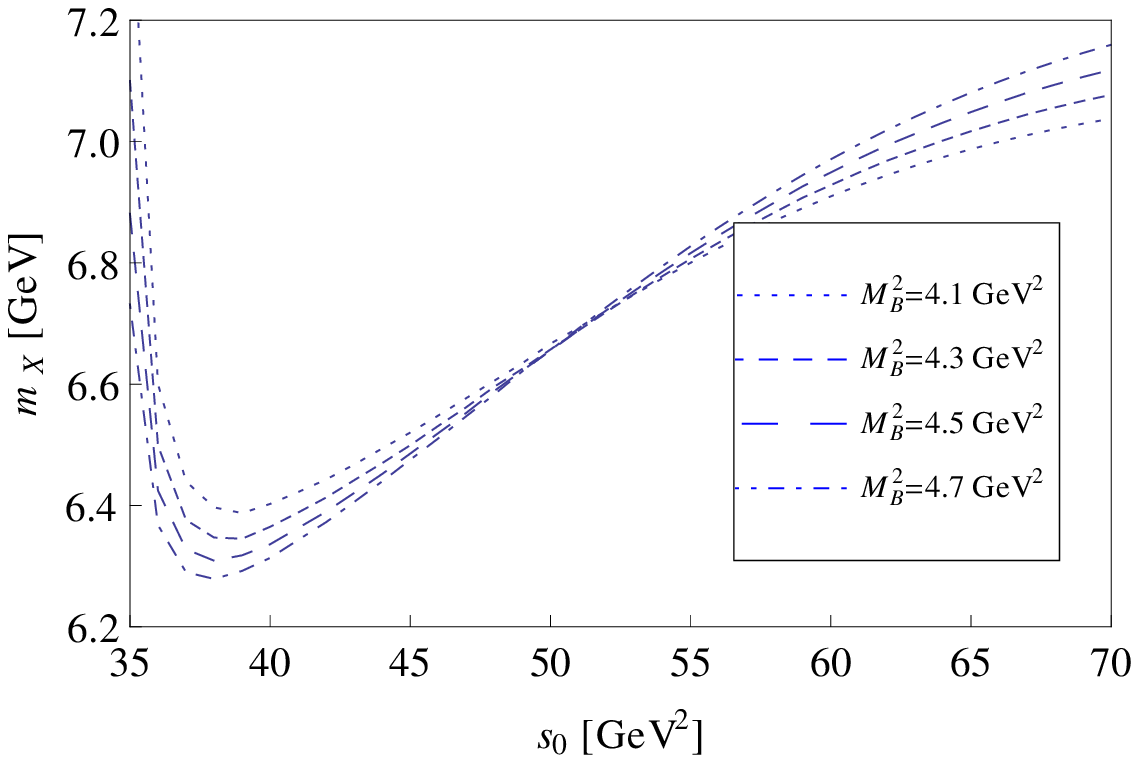}}&
\scalebox{0.65}{\includegraphics{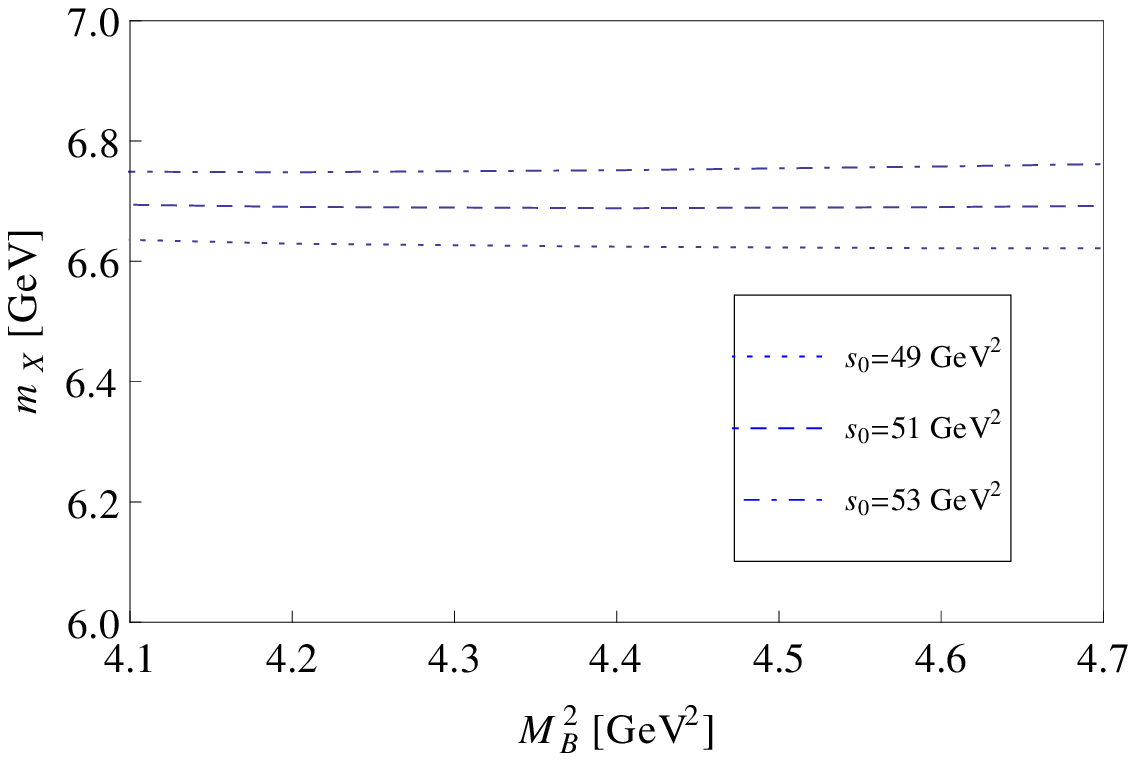}}
\end{tabular}
\figcaption{The variations of the hybrid mass $m_X$ with $s_0$ and $M_B^2$ for $\tilde J^{(2)}_{\mu}$ channel with $J^P=1^-$ adopting Ioffe's condensate values~\cite{2006-Ioffe-p232-277}.} \label{fig1-Ioffe}
\end{center}
%\end{figure}

In the MIT bag model~\cite{1983-Barnes-p241-241, 1983-Chanowitz-p211-211}, the hybrids with $J^{PC}=(0, 1, 2)^{-+}, 1^{--}$ were predicted 
to form the lightest hybrid supermultiplet consisting of a S-wave quark-antiquark pair coupled to an excited gluonic field with 
$J_g^{P_gC_g}=1^{+-}$. A higher hybrid supermultiplet composed of a P-wave $q\bar q$ pair and the same gluonic excitation would contain states with $J^{PC}=0^{+-}, (1^{+-})^3, (2^{+-})^2, 3^{+-}, (0, 1, 2)^{++}$, where the superscript denotes the number of such 
states~\cite{2012-Liu-p126-126, 2011-Dudek-p74023-74023}. These hybrid supermultiplet structures were confirmed for the heavy quark sector in lattice QCD~\cite{2012-Liu-p126-126}, the P-wave quasigluon approach~\cite{2008-Guo-p56003-56003} and QCD sum-rules~\cite{2013-Chen-p19-19} with charmonium and bottomonium interpolating currents in Eq.~\eqref{currents}. 
In Ref.~\cite{2013-Chen-p19-19} the mass of the hybrid with $J^{PC}=0^{--}$ is very high, which may imply a different type of gluonic excitation. 

These supermultiplet structures still exist in the $\bar bGc$ hybrid systems. In Table~\ref{table1}, 
the hybrid states with $J^{P}=(0, 1, 2)^{-}, 1^{-}$ form the lightest supermultiplet while the states with $J^{P}=(0^{+})^2, (1^{+})^2, 2^{+}$ 
form a heavier supermultiplet. The heaviest state is a hybrid with $J^P=0^-$. The two $1^-$ hybrids lie very close since both of them belong to 
the lightest hybrid supermultiplet. In the heavier supermultiplet, there are two $1^+$ hybrids and two $0^+$ hybrids. The mass differences are $0.53$ GeV and $1.18$ GeV for $1^+$ and $0^+$, respectively. The mass difference of about $1.58$ GeV between the two $J^P=0^-$ hybrids is very large, suggesting that the operators are separately probing a ground and excited state. Our interpretation is that these two hybrids have very different gluonic excitations. However, further investigations in other methods are needed to understand the physics of $\bar bGc$ hybrids.

The bottom-charm hybrids are not eigenstates of C-parity and G-parity, so the flavourless hybrids with $J^{PC}=J^{\pm\pm}$ and 
$J^{\pm\mp}$ have the same quantum numbers in the bottom-charm sector and thus can mix. For example, the interpolating currents 
$J^{(1)}_{\mu}$ and $\tilde J^{(2)}_{\mu}$ for the charmonium $\bar cGc$ systems can couple to $1^{-+}$ and $1^{--}$ channels respectively 
and represent totally different channels with opposite C-parity and G-parity. In the MIT bag model~\cite{1983-Barnes-p241-241, 1983-Chanowitz-p211-211}, these two states have different spin configurations of the $\bar qq$ basis. The $1^{--}$ channel has a spin-singlet $S=0$ $\bar qq$ pair while $1^{-+}$ channel contains a spin-triplet $S=1$ $\bar qq$ pair. For the $\bar bGc$ systems, however, they couple to the same $1^-$ bottom-charm channel, so it is not surprising that we obtain two $1^-$ states in Table~\ref{table1}. 

We can also perform a similar sum rule analysis for the coupling constant $f_X$ using Eq.~\eqref{coupling}, in which the hybrid mass $m_X$ can be expressed as in Eq.~\eqref{mass}.
In the coupling sum rules, we use the same criteria as those in the mass sum rules to obtain the working regions of $s_0$ and $M_B^2$. The lower bound on $M_B^2$ determined from 
the OPE convergence is unchanged. Now we perform an analysis of the coupling constant $f_X^2$ as a function of $s_0$. In the left portion of Fig.~\ref{figfX}, we show the variation 
of $f^2_X$ with $s_0$ for the $\tilde J^{(2)}_{\mu}$ channel with $J^P=1^-$. In this figure, the optimized choice of the continuum threshold is $s_0=49$ GeV$^2$, around which the 
variation 
of $f^2_X$ with the Borel mass $M_B$ is minimum. This value is very close to that obtained in the mass sum rules for the same channel. Since the upper bound on $M_B^2$ is only
determined by the value of $s_0$, the Borel window in the coupling sum rules is almost the same as that obtained in the mass sum rules. After performing the numerical analysis, we find that
this situation occurs in all channels. We show the variation of $f^2_X$ with $M_B^2$ for $\tilde J^{(2)}_{\mu}$ channel with $J^P=1^-$ in the right portion of Fig.~\ref{figfX}, 
which demonstrates that the Borel curve is very stable in the working region of Borel mass. For convenience but no loss of generality, we can use the same working regions of $s_0$ 
and $M_B^2$ as those utilized in the mass sum rules to predict the coupling constant $f_X$. The coupling constants for all channels are then calculated and collected in Table~\ref{table1}. The error sources are the same as those in the mass predictions. 
%%%%%%%%%%%%%%%%%%%%%%%%%%%%%%%%%%%%%%%%%%%%%%%%%%%%%%%%%%%%%%%%%%%%%%
%\begin{figure}
\begin{center}
\begin{tabular}{lr}
\scalebox{0.65}{\includegraphics{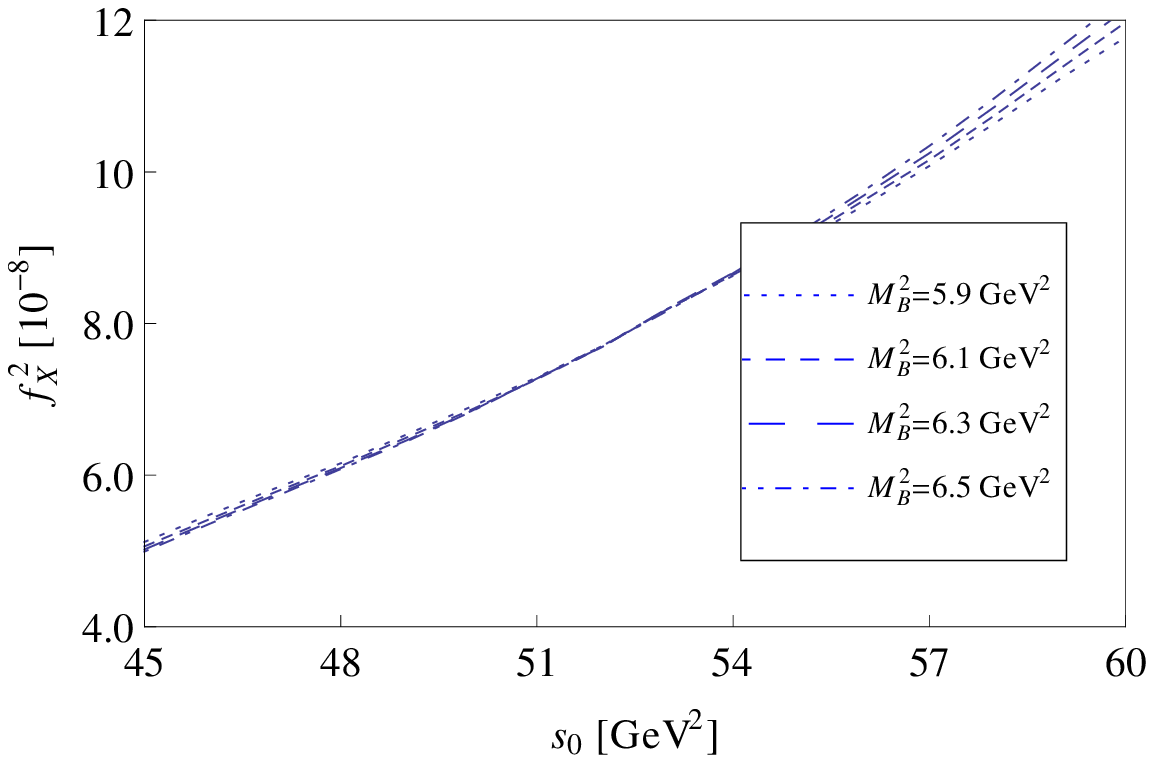}}&
\scalebox{0.65}{\includegraphics{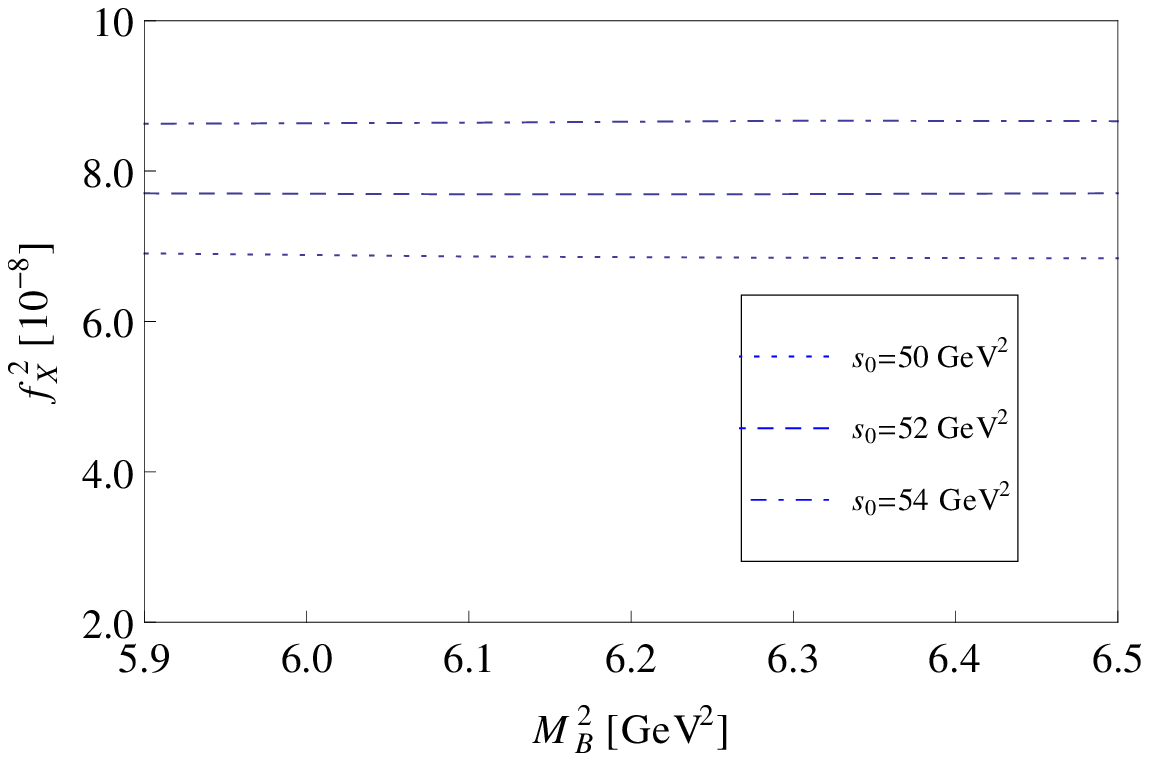}}
\end{tabular}
\figcaption{The variations of $f^2_X$ with $s_0$ and $M_B^2$ for $\tilde J^{(2)}_{\mu}$ channel with $J^P=1^-$.} \label{figfX}
\end{center}
\section{Decay Patterns of the $\bar cGc$, $\bar bGc$ and $\bar bGb$ Hybrids}\label{sec:DECAY}
%==============================================================================================
%==============================================================================================
In this section, we study the decay patterns of the possible $\bar bGc$, $\bar bGb$, and $\bar cGc$ hybrid states 
% \trem{For comparison, we also study the possible 
% decay modes of the charmonium and bottomonium hybrids, which have been studied in}
using the mass predictions obtained in Table \ref{table1} and 
 Ref.~\cite{2013-Chen-p19-19}. We just consider the 
two-body hadronic decays. Both the open flavour and hidden flavour deacy modes are taken into account.

According to some model-dependent analyses in Refs.~\cite{1985-Isgur-p2910-2910,1985-Isgur-p869-869,1995-Close-p233-254,2005-Kou-p164-169,
1999-Zhu-p14008-14008,2005-Close-p215-222}, hybrid mesons prefer to decay into $S+P$-wave final states. For example, the decay modes to 
pairs of identical $S$-wave mesons might be suppressed as shown in Refs.~\cite{2005-Kou-p164-169,1999-Zhu-p14008-14008,2005-Close-p215-222}.
For the charmonium hybrids, this implies that the $D\bar D, D^*\bar D^*$ and $D_s\bar D_s, D^*_s\bar D^*_s$ decay modes are suppressed 
while $D^*\bar D$ and $D^*_s\bar D_s$ are very small. The $D^{(*)}\bar D^{*}_0$ and $D^{(*)}\bar D_1$  channels would be the dominant decay modes if 
the hybrids are above their thresholds. In the MIT bag model~\cite{1983-Barnes-p241-241, 1983-Chanowitz-p211-211}, the $q\bar q$ pairs in hybrid 
mesons with $J^{PC}=1^{--}, 0^{++}, 1^{++}, 2^{++}$ are in a net spin singlet configuration. These hybrid mesons are forbidden to decay into 
final states consisting only of spin singlet mesons due to the spin selection rule~\cite{1999-Page-p34016-34016}.

Using the masses predicted in Ref.~\cite{2013-Chen-p19-19}, we collect the possible $S$-wave and $P$-wave decay modes of the charmonium hybrids 
in Table~\ref{cGc decay} by taking account into the conservation of quantum numbers and the selection rules mentioned above. All the open charm 
decay modes should be understood as containing the charge conjugation parts. We have not listed the $D^{(*)}_s\bar D^{(*)}_s$ decay modes in 
Table~\ref{cGc decay} because they are similar to the $D^{(*)}\bar D^{(*)}$ channels. One notes that the charmonium hybrids with 
$J^{PC}=1^{--}, 0^{-+}, 1^{-+}$, which belong to the lightest supermultiplet, lie below the open charm $D^{(*)}\bar D^{(*)}$ thresholds. 
They just decay via the hidden charm mechanism. The positive-parity states with $J^{PC}=1^{+-}, (0, 1, 2)^{++}$ belong to the heavier 
hybrid supermultiplet. They lie above the open charm thresholds. According to the $S+P$-wave selection rule, they prefer the $D^{(*)}\bar D^{*}_0$ 
and $D^{(*)}\bar D_1$ decays. In Table~\ref{cGc decay}, these decay modes are the $P$-wave coupling channels. These features are very 
different from the conventional $c\bar c$ mesons and the other exotic configurations such as the tetraquarks $cq\bar c\bar q$ and molecules 
$c\bar q\bar cq$, for which the $S$-wave $D\bar D, D^*\bar D^*, D\bar D^*$ channels are favoured. If the charmonium hybrids $\bar cGc$ are 
predicted above the $D^{(*)}\bar D^{*}_0$ and $D^{(*)}\bar D_1$ thresholds, the observation of the anomalous branching ratios in these different 
channels could be understood as a strong hybrid signature~\cite{1995-Close-p233-254,1995-Barnes-p5242-5256,2005-Close-p215-222}.

%%%%%%%%%%%%%%%%%%%%%%%%%%%%%%%%%%%%%%%%%%%%%%%%%%%%%%%%%%%%%%%%%%%%%%%%%%%%%%%%%%%%%%%%%%%%%%%%%%%%%%%%%%%%%%%%%%%%%%%%%%%%%%%%%%%%%%%
\begin{center}
\renewcommand{\arraystretch}{1.3}
\begin{tabular*}{16cm}{ccc}
%\begin{tabular}{ccc}
\hlinewd{.8pt}
$I^GJ^{PC}$  &                     $S$-wave                                            &                           $P$-wave                \\
\hline
$0^-1^{--}$  & $-$                                                                     & $-$ \\
$0^+0^{-+}$  & $\eta_c(1S)f_0(600)$                                                    & $-$\\
$0^+1^{-+}$  & $-$                                                                     & $\eta_c(1S)\eta, J/\psi\omega(782)$\\
$0^+2^{-+}$  & $-$                                                                     & $D\bar D^*, J/\psi\omega(782)$
\vspace{10pt}\\
$0^-0^{+-}$  & $-$                                                                     & $J/\psi f_0(600)$\\
$0^+2^{++}$  & $J/\psi\omega(782), J/\psi\phi(1020),$                                  & $D\bar D_1, D\bar D_2^*, D^*\bar D_0^*, D^*\bar D_1, D^*\bar D_0^*,$\\
             & $\chi_{c2}(1P)f_0(600)$                                                 & $J/\psi h_1(1170), \eta_c(1S)f_1(1285), \eta_c(1S)f_2(1270),$\\
             &                                                                         & $\chi_{c1}(1P)\eta, \chi_{c2}(1P)\eta$\\
$0^-1^{+-}$  & $D\bar D^*, J/\psi\eta, \psi(2S)\eta, \chi_{c0}(1P)h_1(1170)$           & $D\bar D_0^*, D\bar D_1, D\bar D_2^*, D^*\bar D_0^*, D^*\bar D_2^*, D^*\bar D_1,$\\
             &                                                                         & $\eta_c(1S)h_1(1170), \eta_c(2S)h_1(1170)$\\
$0^+1^{++}$  & $D\bar D^*, D^*_0\bar D_1, D_1\bar D_2^*, J/\psi\omega(782),$           & $D\bar D_0^*, D\bar D_1, D\bar D_2^*, D^*\bar D_0^*, D^*\bar D_2^*, D^*\bar D_1,$\\
             & $J/\psi\phi(1020),\chi_{c0}(1P)f_1(1285),$                              & $\eta_c(1S)f_0(600), \eta_c(1S)f_0(980),\eta_c(2S)f_0(600),$\\
             & $\chi_{c1}(1P)f_0(600), \chi_{c2}(1P)f_1(1285)$                         & $\eta_c(2S)f_0(980), \eta_c(1S)f_1(1285), \eta_c(1S)f_2(1270),$\\
             &                                                                         & $ J/\psi h_1(1170), \chi_{c(0,1,2)}(1P)\eta$\\
$0^+0^{++}$  & $J/\psi\omega(782), J/\psi\phi(1020), \eta_c(1S)\eta, \eta_c(2S)\eta,$  & $D\bar D_1, D^*\bar D_0^*, D^*\bar D_1, D^*\bar D_2^*,$\\
             & $\chi_{c0}(1P)f_0(600), \chi_{c0}(1P)f_0(980),$                         & $\eta_c(1S)f_1(1285), \eta_c(2S)f_1(1285), J/\psi h_1(1170), $\\
             & $\chi_{c1}(1P)f_1(1285), \chi_{c2}(1P)f_2(1270)$                        & $\psi(2S)h_1(1170), \chi_{c1}(1P)\eta$
\vspace{10pt}\\
$0^-0^{--}$  & $D\bar D_0^*, D^*\bar D_1, J/\psi f_1(1285),$                           & $D\bar D^*, D_0^*\bar D_1, D_1\bar D_2^*, J/\psi\eta, \psi(2S)\eta,$\\
             & $\psi(2S)f_1(1285), \chi_{c1}(1P)\omega(782)$                           & $\eta_c(1S)\omega(782), \eta_c(2S)\omega(782), \chi_{c(0,1,2)}(1P)h_1(1170)$ \\
\hlinewd{.8pt}
%\end{tabular}
\end{tabular*}
\tabcaption{The possible decay modes of the charmonium hybrids with various quantum numbers. \label{cGc decay}}
\end{center}
%%%%%%%%%%%%%%%%%%%%%%%%%%%%%%%%%%%%%%%%%%%%%%%%%%%%%%%%%%%%%%%%%%%%%%%%%%%%%%%%%%%%%%%%%%%%%%%%%%%%%%%%%%%%%%%%%%%%%%%%

Replacing the D and charmonium mesons by the B and bottomonium mesons respectively, we obtain the decay patterns of the bottomonium hybrids 
in Table~\ref{bGb decay} so long as the kinematics allows. We note that the masses of the bottomonium hybrids with 
$J^{PC}=1^{--}, 0^{-+}, 1^{-+}, 2^{-+}$ in the lightest supermultiplet are lower than the open bottom $B^{(*)}\bar B^{(*)}$, 
$B_s{(*)}\bar B_s^{(*)}$ thresholds and one bottomonium meson plus one light meson thresholds. They cannot decay into the two-body final states 
through neither  open bottom nor hidden bottom mechanisms. Once produced, they only decay via the electromagnetic and weak interactions.

The decay mechanism of the $\bar bGc$ hybrids will be different from the $\bar cGc$ and $\bar bGb$ states. The selection rule forbidding $S+S$-wave 
final states no longer works in this situation because the internal structures and sizes of the $D$ and $B$ mesons differ~\cite{1995-Close-p233-254,1995-Close-p1706-1709,1997-Swanson-p5692-5695}. The $B^{(*)}D^{(*)}$ and $B_s^{(*)}D_s^{(*)}$ open flavour decay 
modes are prefered for the $\bar bGc$ hybrids above these thresholds. Besides, there will be no constraints of the C-parity and G-parity as for 
the $\bar cGc$ and $\bar bGb$ hybrids since the $\bar bGc$ hybrids are not enginstates of C-parity and G-parity. We collect the possible 
decay patterns of the $\bar bGc$ hybrids in Table~\ref{bGc decay}. 
The PDG mass of the pesudoscalar $B_c$ meson is $m_{B_c}=6.277\pm0.006$ GeV~\cite{2012-Beringer-p10001-10001}. To date, the other bottom-charm $B_c$ 
mesons have not been observed. To predict the hidden flavour decays of the $\bar bGc$ hybrids, we use the masses of the vector ($B_c^*$), scalar ($B_{c0}$), 
axialvector ($B_{c1}$) and tensor ($B_{c2}$) bottom-charm mesons predicted in lattice QCD~\cite{1996-Davies-p131-137}: $m_{B_c^*}=6.321$ GeV, $m_{B_{c0}}=6.727$ GeV, 
$m_{B_{c1}}=6.743$ GeV and $m_{B_{c2}}=6.783$ GeV. In contrast to the $\bar cGc$ and $\bar bGb$ hybrids, more decay modes are alowed and the $S+S$-wave pair 
decays are dominant.
%%%%%%%%%%%%%%%%%%%%%%%%%%%%%%%%%%%%%%%%%%%%%%%%%%%%%%%%%%%%%%%%%%%%%%%%%%%%%%%%%%%%%%%%%%%%%%%%%%%%%%%%%%%%%%%%%%%%%%%%%%%%%%%%%%%%%%%
\begin{center}
\renewcommand{\arraystretch}{1.25}
\begin{tabular*}{15.3cm}{ccc}
%\begin{tabular}{ccc}
\hlinewd{.8pt}
$I^GJ^{PC}$  &                     $S$-wave                                            &                           $P$-wave                \\
\hline
$0^-1^{--}$  & $-$                                                                     & $-$ \\
$0^+0^{-+}$  & $-$                                                                     & $-$\\
$0^+1^{-+}$  & $-$                                                                     & $-$\\
$0^+2^{-+}$  & $-$                                                                     & $-$
\vspace{10pt}\\
$0^-0^{+-}$  & $-$                                                                     & $\Upsilon(1S)f_0(600)$\\
$0^+2^{++}$  & $\Upsilon(1S)\omega(782), \Upsilon(1S)\phi(1020), \chi_{b2}(1P)f_0(600)$& $\Upsilon(1S)h_1(1170), \chi_{b1}(1P)\eta, \chi_{b2}(1P)\eta$\\
$0^-1^{+-}$  & $B\bar B^*, \Upsilon(1S)\eta, \Upsilon(2S)\eta$                         & $-$\\
$0^+1^{++}$  & $B\bar B^*,$                                                            & $B\bar B_1, B\bar B_2^*, B^*\bar B_2^*, B^*\bar B_1,$\\
             & $\Upsilon(1S)\omega(782),\Upsilon(1S)\phi(1020), \chi_{b1}(1P)f_0(600)$ & $\Upsilon(1S)h_1(1170), \chi_{b(0,1,2)}(1P)\eta$\\
$0^+0^{++}$  & $\Upsilon(1S)\omega(782), \Upsilon(1S)\phi(1020),\chi_{b0}(1P)f_0(600),$& $B\bar B_1, B^*\bar B_1, B^*\bar B_2^*,$\\
             & $ \chi_{b0}(1P)f_0(980),\chi_{b1}(1P)f_1(1285),$                        & $\Upsilon(1S)h_1(1170), \Upsilon(2S)h_1(1170), \chi_{b1}(1P)\eta$\\
             & $\chi_{b2}(1P)f_2(1270)$                                                &
\vspace{10pt}\\
$0^-0^{--}$  & $B^*\bar B_1,$                                                          & $B\bar B^*, B_1\bar B_2^*,$\\
             & $\Upsilon(1S)f_1(1285), \Upsilon(2S)f_1(1285), \chi_{b1}(1P)\omega(782)$& $\Upsilon(1S)\eta, \Upsilon(2S)\eta, \chi_{b(0,1,2)}(1P)h_1(1170)$ \\
\hlinewd{.8pt}
%\end{tabular}
\end{tabular*}
\tabcaption{The possible decay modes of the bottomonium hybrids with various quantum numbers. \label{bGb decay}}
\end{center}
%%%%%%%%%%%%%%%%%%%%%%%%%%%%%%%%%%%%%%%%%%%%%%%%%%%%%%%%%%%%%%%%%%%%%%%%%%%%%%%%%%%%%%%%%%%%%%%%%%%%%%%%%%%%%%%%%%%%%%%%

%%%%%%%%%%%%%%%%%%%%%%%%%%%%%%%%%%%%%%%%%%%%%%%%%%%%%%%%%%%%%%%%%%%%%%%%%%%%%%%%%%%%%%%%%%%%%%%%%%%%%%%%%%%%%%%%%%%%%%%%%%%%%%%%%%%%%%%
\begin{center}
\renewcommand{\arraystretch}{1.25}
\begin{tabular*}{15.3cm}{ccc}
\hlinewd{.8pt}
$I(J^{P})$  &                     $S$-wave                                            &                           $P$-wave                \\
\hline
$0(1^{-})$  & $-$                                                                     & $-$ \\
$0(0^{-})$  & $-$                                                                     & $-$\\
$0(1^{-})$  & $-$                                                                     & $B_c\eta, B_c^*\eta$\\
$0(2^{-})$  & $-$                                                                     & $B_c\omega(782), B_c^*\eta$
\vspace{10pt}\\
$0(0^{+})$  & $BD, B^*D^*, B_c\eta, B_c^*\omega(782), B_{c0}f_0(600)$                 & $B_c^*f_0(600)$\\
$0(2^{+})$  & $B^*D^*,$                                                               & $B_cf_1(1285), B_cf_2(1270), B_c^*f_0(600),$\\
            & $B_c^*\omega(782), B_c^*\phi(1020), B_{c2}f_0(600)$                     & $B_c^*f_0(980), B_c^*h_1(1170), B_{c1}\eta, B_{c2}\eta$\\
$0(1^{+})$  & $BD^*, B^*D^*, B_c\omega(782), B_c^*\eta,$                              & $BD_0^*, BD_1, BD_2^*, B^*D_0^*, B_cf_0(600), B_cf_0(980),$\\
            & $B_c^*\omega(782), B_c^*\phi(1020), B_{c1}f_0(600)$                     & $B_ch_1(1170), B_cf_1(1285), B_c^*f_0(600), B_c^*f_0(980)$\\
$0(1^{+})$  & $BD^*, B^*D^*, B_1D^*_0, B_1D_2^*,$                                     & $BD_0^*, BD_1, BD_2^*, B^*D_0^*, B^*D_2^*, B^*D_1,$\\
            & $B_c\omega(782), B_c^*\eta, B_c\phi(1020), B_c^*\omega(782),$           & $B_cf_0(600), B_cf_0(980), B_ch_1(1170),$\\
            & $B_{c(0,1,2)}f_1(1285), B_c^*\phi(1020), B_{c1}f_0(600)$                & $B_cf_1(1285), B_c^*f_0(600), B_c^*f_0(980),$\\
            &                                                                         & $B_{c(0,1,2)}\omega(782), B_{c(0,1,2)}\phi(1020)$\\
$0(0^{+})$  & $BD, B^*D^*, B_1D_1, B_c\eta, B_c^*\omega(782),$                        & $BD_1, B^*D_0^*, B^*D_1, B^*D_2^*,$\\
            & $B_c^*\phi(1020), B_{c0}f_0(600), B_{c0}f_0(980),$                      & $B_cf_1(1285), B_c^*f(600), B_c^*f(980), B_c^*h_1(1170), $\\
            & $B_{c1}h_1(1170), B_{c1}f_1(1285), B_{c2}f_2(1270)$                     & $B_c^*f_1(1285), B_{c1}\eta, B_{c1}\omega(782), B_{c1}\phi(1020)$
\vspace{10pt}\\
$0(0^{-})$  & $BD_0^*, B^*D_1,$                                                       & $BD^*, B^*D^*, B_1D_0^*, B_1D_1, B_1D_2^*,$\\
            & $B_cf_0(600), B_cf_0(980), B_{c0}\eta,$                                 & $B_c\omega(782), B_c\phi(1020), B_c^*\eta,$ \\
            & $B_c^*f_1(1285), B_{c1}\omega(782), B_{c1}\phi(1020)$                   & $B_{c(0,1,2)}h_1(1170), B_{c(0,1,2)}f_1(1285)$\\
\hlinewd{.8pt}
\end{tabular*}
\tabcaption{The possible decay modes of the $\bar bGc$ hybrids with various quantum numbers. \label{bGc decay}}
\end{center}
%%%%%%%%%%%%%%%%%%%%%%%%%%%%%%%%%%%%%%%%%%%%%%%%%%%%%%%%%%%%%%%%%%%%%%%

%================================================================================
%================================================================================
\section{SUMMARY}\label{sec:SUMMARY}
%================================================================================
%================================================================================
In this paper we have studied the $\bar bGc$ hybrid systems in QCD sum rules using the interpolating currents in Eq.~\eqref{currents}.
We have calculated the correlation functions and the spectral densities up to dimesion six at leading order in $\alpha_s$. 
After performing the QCD sum rule analysis, we have extracted the masses of the possible $\bar bGc$ hybrid states with 
$J^P=0^{-}, 0^{+}, 1^{-}, 1^{+}, 2^{-}, 2^{+}$.

We have calculated the perturbative terms, gluon condensate 
and the dimension six tri-gluon condensate contributions. The gluon condensate is the dominant power correction to the correlation functions. 
However, the tri-gluon condensate is also very important since it can stablize the hybrid sum rules as found in Refs.~\cite{2012-Qiao-p15005-15005,2012-Harnett-p125003-125003,2012-Berg-p34002-34002,2013-Chen-p19-19}. Since the $\bar bGc$ hybrids are charged 
states, they don't have definite C-parities. For the quantum numbers $J^P=0^-, 0^+, 1^-, 1^+$, there are two interpolating currents 
in Eq.~\eqref{currents} that could couple to them. One should not expect identical results from the two currents since they are simply 
probes of the hadronic spectrum, and may have different couplings to the ground and excited states. In Table~\ref{table1}, we extract 
two different masses for each quantum numbers with $J^P=0^-, 0^+, 1^-, 1^+$. The two $1^-$ hybrids lie very close to each other since 
both of them belong to the lightest hybrid supermultiplet. Although they have the same gluonic excitations and orbital excitations 
between $\bar b$ and $c$ quarks, the spin configurations of the $\bar bc$ pair are different~\cite{1983-Barnes-p241-241, 1983-Chanowitz-p211-211}. 
The $1^-$ hybrid extracted from $J_1^{\mu}(x)$ is a spin-triplet state while the other is a spin-triplet state.
The mass differences between two $J^P=1^+$ hybrids and two $J^P=0^+$ hybrids are much bigger. These four hybrids belong to the heavier supermultiplet. 
The biggest mass difference occurs for the two $J^P=0^-$ states. One of them belongs to the lightest hybrid supermultiplet while another one may have 
a very different excitation of the gluonic field since it appears at a much higher mass scale.

We have also predicted the possible open-flavour and hidden-flavour decay patterns of the $\bar cGc$, $\bar bGc$ and $\bar bGb$ hybrids. 
If the $S+P$-wave selection rule turns out to be correct, the $S+P$-wave final states such as $D^{(*)}_{(s)}\bar D^{*}_{0(s)}, D^{(*)}_{(s)}\bar D_{1(s)}$ 
are dominant decay modes for the $\bar cGc$ and $\bar bGb$ hybrids. For the $\bar bGc$ hybrids, however, the $S+P$-wave selection rule is not in operation 
so that the most important decay modes are $B^{(*)}D^{(*)}$ and $B_s^{(*)}D_s^{(*)}$.

To our knowledge, the $\bar bGc$ systems have not been studied before, and thus our work provides important benchmarks for future investigations of the $\bar bGc$ hybrids in other phenomenological methods. Hopefully our investigation in this work will be useful to the future search of these states at the experimental facilities such as BESIII, PANDA and LHCb.
%%%%%%%%%%%%%%%%%%%%%%%%%%%%%%%%
\section*{Acknowledgments}
%%%%%%%%%%%%%%%%%%%%%%%%%%%%%%%%

This project was supported by the Natural Sciences and Engineering
Research Council of Canada (NSERC). S.L.Z. was supported by the
National Natural Science Foundation of China under Grants
11075004, 11021092, 11261130311 and Ministry of Science and
Technology of China (2009CB825200).

%\bibliographystyle{h-physrev}

%\bibliography{myreference}

%%%%%%%%%%%%%%%%%%%%%%%%%%%%%%%%%%%%%%%%%%%%%%%%%%%%%%%%%%%%%%%%%%%%
\appendix

%=====================================================================================
%=====================================================================================
\section{Spectral Densities}\label{sec:rhos}
%=====================================================================================
%=====================================================================================
In this appendix, we list the spectral densities of the hybrid interpolating currents in Eq.~(\ref{currents}). We calculate the spectral densities up to dimension six:
\begin{eqnarray}
\rho(s)=\rho^{pert}(s)+\rho^{\GGa}(s)+\rho^{\GGGa}(s),
\end{eqnarray}

For all of the various $J^{P}$ quantum numbers studied in this paper, the perturbative contributions can be expressed as
\begin{align}
\nonumber\rho^{pert}(s)&=-\dab\frac{\alpha_s(1-\alpha-\beta)(m_1^2\beta+m_2^2\alpha-\alpha\beta s)}{24\pi^3\alpha^2\beta^3}\times\\
&\left[f_1(\alpha,\beta)(m_1^2\beta+m_2^2\alpha)^2+f_2(\alpha,\beta)(m_1^2\beta+m_2^2\alpha)\alpha\beta s+f_3(\alpha,\beta)m_1m_2
+f_4(\alpha,\beta)(\alpha\beta s)^2\right], \label{perturbative contribution}
\end{align}
where $\alpha_{min}=\frac{1}{2}\left\{1+\frac{m_1^2-m_2^2}{s}-\left[\left(1+\frac{m_1^2-m_2^2}{s}\right)^2-\frac{4m_1^2}{s}\right]^{1/2}\right\}$, $\alpha_{max}=\frac{1}{2}\left\{1+\frac{m_1^2-m_2^2}{s}+\left[\left(1+\frac{m_1^2-m_2^2}{s}\right)^2-\frac{4m_1^2}{s}\right]^{1/2}\right\}$, 
$\beta_{min}=\frac{\alpha m_2^2}{\alpha s-m_1^2}$, $\beta_{max}=1-\alpha$. $m_1$ and $m_2$ are the heavy quark masses.  
$f_1(\alpha,\beta)$, $f_2(\alpha,\beta)$, $f_3(\alpha,\beta)$ and $f_4(\alpha,\beta)$ are polynomials in $\alpha$ and $\beta$. 
For the dimension four gluon condensate $\GGb$, all contributions can be expressed as
\begin{align}
\rho^{\GGa}(s)&=\frac{\GGb}{144\pi}\left\{g_1(s)\left[\left(1-\frac{m_1^2-m_2^2}{s}\right)^2-\frac{4m_2^2}{s}\right]^{3/2}+\left[g_2(s)+g_3(s)m_1m_2\right]\left[\left(1-\frac{m_1^2-m_2^2}{s}\right)^2-\frac{4m_2^2}{s}\right]^{1/2}\right\}, \label{gluon condensate contribution}
\end{align}
where $g_1(s), g_2(s), g_3(s)$ are polynomials in $s$. The spectral densities of the dimension six tri-gluon condensate $\GGGb$ can be written as
\begin{align}
\nonumber\rho^{\GGGa}(s)=-\frac{\GGGb}{192\pi^2}\int^1_0dx\Big\{&\left[h_1(x)+h_2(x)m_1m_2\right]\delta'(s-\tilde{m}^2)
+\left[h_3(x)+h_4(x)m_1m_2\right]\delta(s-\tilde{m}^2)
\\&+h_5(x)\theta(s-\tilde{m}^2)\Big\}, \label{tri-gluon condensate contribution}
\end{align}
in which $\delta'(s-\tilde m^2)=\frac{\partial \delta(s-\tilde m^2)}{\partial s}$, $\tilde m^2=\frac{m_1^2x+m_2^2(1-x)}{x(1-x)}$ where $x$ is a Feynman parameter.
$h_1(x)$, $h_2(x)$, $h_3(x)$, $h_4(x)$, $h_5(x)$ are polynomials in $x$.

We tabulate the polynomials $f_1(\alpha,\beta)$, $f_2(\alpha,\beta)$, $f_3(\alpha,\beta)$, $f_4(\alpha,\beta)$ for the perturbative contribution~\eqref{perturbative contribution},  $g_1(s), g_2(s), g_3(s)$ for the gluon condensate contribution~\eqref{gluon condensate contribution} and $h_1(x)$, $h_2(x)$, $h_3(x)$, $h_4(x)$, $h_5(x)$ for tri-gluon condensate 
contribution~\eqref{tri-gluon condensate contribution} for $J=0, 1, 2$ in Table~\ref{polynomials} (up to a sign). The signs of these polynomials are specified in 
Table~\ref{signs of polynomials} for various $J^P$ quantum numbers. Note that the tri-gluon condensate contribution~\eqref{tri-gluon condensate contribution} contains $\delta(s-\tilde m^2)$ and its derivatives. They should be there to compensate for the singular behavior of the spectral densities at threshold  $s=(m_1+m_2)^2$. 
We keep the integral forms of $\rho^{\GGGa}$ in our sum rule analyses.

%%%%%%%%%%%%%%%%%%%%%%%%%%%%%%%%%%%%%%%%%%%%%%%%%%%%%%%%%%%%%%%%%%%%
\begin{center}
\renewcommand{\arraystretch}{1.3}
\begin{tabular*}{17.5cm}{ccccc}
%\begin{tabular}{ccccc}
\hlinewd{.8pt}
&$J$ & $0$ & $1$ & $2$\\
\hline
&$f_1(\alpha,\beta)$ &$9\alpha-9\alpha^2+4\beta-12\alpha\beta-3\beta^2$&$-9\alpha+9\alpha^2-4\beta+12\alpha\beta+3\beta^2$                                  
                                                                                             & $\frac{1}{2}(-6\alpha+6\alpha^2-3\beta+7\alpha\beta+\beta^2)$\\
&$f_2(\alpha,\beta)$ &$9-36\alpha+27\alpha^2-$           & $-(15-66\alpha+51\alpha^2-$       & $-\frac{1}{2}(12-54\alpha+42\alpha^2-$\\
                    &&$41\beta+60\alpha\beta+33\beta^2$  & $29\beta+60\alpha\beta+9\beta^2)$ & $27\beta+47\alpha\beta+5\beta^2)$\\ 
&$f_3(\alpha,\beta)$ &$3(\alpha\beta+\beta^2)(m_1^2\beta+m_2^2\alpha-\alpha\beta s)+$        &$-3(\alpha\beta+\beta^2)(m_1^2\beta+m_2^2\alpha-\alpha\beta s)$&$0$\\
                    &&$12(1-\alpha-\beta)\alpha\beta^2 s$ & &\\
&$f_4(\alpha,\beta)$ &$-(9-27\alpha+18\alpha^2-$         &$27-81\alpha+54\alpha^2-$          &$2(6-18\alpha+12\alpha^2-$\\
                    &&$49\beta+60\alpha\beta+42\beta^2)$ &$37\beta+60\alpha\beta+6\beta^2$   &$9\beta+13\alpha\beta+\beta^2)$
\vspace{8pt}\\
&$g_1(s)$&$3s$& $s$ & $0$\\
&$g_2(s)$&$-9(s-m_1^2-m_2^2)$& $3(s-m_1^2-m_2^2)$ &$0$\\
&$g_3(s)$& $-18$& $18$ &$6$
\vspace{8pt}\\
&$h_1(x)$&$\frac{2x(3-8x+4x^2)m_1^4}{(1-x)^4}-\frac{2(6x-5)m_1^2m_2^2}{(1-x)^2}+\frac{4m_2^4}{x}$ &$\frac{2x(2x-1)m_1^4}{(1-x)^4}-\frac{2m_1^2m_2^2}{(1-x)^2}$   &$0$  \\
&$h_2(x)$&$\frac{6(1-2x)m_1^2}{(1-x)^3}+\frac{6m_2^2}{x(1-x)}$ &$\frac{6(2x-1)m_1^2}{(1-x)^3}-\frac{6m_2^2}{x(1-x)}$   &$-\frac{2m_1^2}{(1-x)^2}$ \\
&$h_3(x)$&$\frac{(-3+19x-20x^2+8x^3)m_1^2}{(1-x)^3}+\frac{2(5x-1)m_2^2}{x}$ &$\frac{(1-7x+6x^2-4x^3)m_1^2}{(1-x)^3}-2m_2^2$   &$0$ \\
&$h_4(x)$&$\frac{3(3x-1)}{x(1-x)^2}$ &$\frac{3(1-3x)}{x(1-x)^2}$   &$0$ \\
&$h_5(x)$&$(6x-2)$ &$-(6x-2)$   &$0$ \\
\hline
\hlinewd{.8pt}
%\end{tabular}
\end{tabular*}
\tabcaption{Polynomials $f_1(\alpha,\beta)$, $f_2(\alpha,\beta)$, $f_3(\alpha,\beta)$, $f_4(\alpha,\beta)$ for the perturbative contribution~\eqref{perturbative contribution},  $g_1(s), g_2(s), g_3(s)$ for the gluon condensate contribution~\eqref{gluon condensate contribution} and $h_1(x)$, $h_2(x)$, $h_3(x)$, $h_4(x)$, $h_5(x)$ for the tri-gluon condensate contribution~\eqref{tri-gluon condensate contribution} for $J=0, 1, 2$.} \label{polynomials}
\end{center}
%%%%%%%%%%%%%%%%%%%%%%%%%%%%%%%%%%%%%%%%%%%%%%%%%%%%%%%%%%%%%%%%%%%%%%%%%%%%%%%%%%%%%%%%%%%%%%%
%%%%%%%%%%%%%%%%%%%%%%%%%%%%%%%%%%%%%%%%%%%%%%%%%%%%%%%%%%%%%%%%%%%%
\begin{center}
\renewcommand{\arraystretch}{1.3}
\begin{tabular*}{15cm}{ccccccccccccccc}
%\begin{tabular}{ccccccccccccccc}
\hlinewd{.8pt} 
&Current&$J^P$  &$f_1(\alpha,\beta)$&$f_2(\alpha,\beta)$&$f_3(\alpha,\beta)$&$f_4(\alpha,\beta)$&$g_1(s)$&$g_2(s)$&$g_3(s)$&$h_1(x)$&$h_2(x)$&$h_3(x)$&$h_4(x)$&$h_5(x)$\\
\hline
&$J^{(1)}_{\mu}$           & $0^{+}$&$+$&$+$&$+$&$+$&$+$&$+$&$+$&$+$&$+$&$+$&$+$&$+$\\
&$J^{(2)}_{\mu}$           & $0^{-}$&$+$&$+$&$-$&$+$&$+$&$+$&$-$&$+$&$-$&$+$&$-$&$+$\\
&$\tilde J^{(1)}_{\mu}$    & $0^{-}$&$+$&$+$&$+$&$+$&$-$&$-$&$-$&$-$&$-$&$-$&$-$&$-$\\
&$\tilde J^{(2)}_{\mu}$    & $0^{+}$&$+$&$+$&$-$&$+$&$-$&$-$&$+$&$-$&$+$&$-$&$+$&$-$
\vspace{8pt}\\
&$J^{(1)}_{\mu}$           & $1^{-}$&$+$&$+$&$+$&$+$&$+$&$+$&$+$&$+$&$+$&$+$&$+$&$+$\\
&$J^{(2)}_{\mu}$           & $1^{+}$&$+$&$+$&$-$&$+$&$+$&$+$&$-$&$+$&$-$&$+$&$-$&$+$\\
&$\tilde J^{(1)}_{\mu}$    & $1^{+}$&$+$&$+$&$+$&$+$&$-$&$-$&$-$&$-$&$-$&$-$&$-$&$-$\\
&$\tilde J^{(2)}_{\mu}$    & $1^{-}$&$+$&$+$&$-$&$+$&$-$&$-$&$+$&$-$&$+$&$-$&$+$&$-$
\vspace{8pt}\\
&$J^{(3)}_{\mu\nu}$        & $2^{-}$&$+$&$+$&$+$&$+$&$+$&$+$&$+$&$+$&$+$&$+$&$+$&$+$\\
&$\tilde J^{(3)}_{\mu\nu}$ & $2^{+}$&$+$&$+$&$+$&$+$&$-$&$-$&$-$&$-$&$-$&$-$&$-$&$-$\\
\hline
\hlinewd{.8pt}
%\end{tabular}
\end{tabular*}
\tabcaption{The signs of the polynomials $f_1(\alpha,\beta)$, $f_2(\alpha,\beta)$, $f_3(\alpha,\beta)$, $f_4(\alpha,\beta)$ for the perturbative contribution~\eqref{perturbative contribution},  $g_1(s), g_2(s), g_3(s)$ for the gluon condensate contribution~\eqref{gluon condensate contribution} and $h_1(x)$, $h_2(x)$, $h_3(x)$, $h_4(x)$, $h_5(x)$ for the tri-gluon condensate contribution~\eqref{tri-gluon condensate contribution} for various $J^P$ quantum numbers.  \label{signs of polynomials}}
\end{center}
%%%%%%%%%%%%%%%%%%%%%%%%%%%%%%%%%%%%%%%%%%%%%%%%%%%%%%%%%%%%%%%%%%%%%%%

\end{document}